\documentclass[12pt,preprint]{aastex}

\shorttitle{Outflow Driven Cavities}
\shortauthors{Cunningham, Frank, Quillen \& Blackman}

\begin{document}

\title{Outflow Driven Cavities: Numerical Simulations of
Intermediaries of Protostellar Turbulence}

\author{Andrew J. Cunningham\altaffilmark{1}, Adam Frank\altaffilmark{2},
 Alice C. Quillen\altaffilmark{3} \& Eric G. Blackman\altaffilmark{4}}
\affil{Department of Physics and Astronomy, University of Rochester,
    Rochester, NY 14620}

\altaffiltext{1}{ajc4@pas.rochester.edu}
\altaffiltext{2}{afrank@pas.rochester.edu}
\altaffiltext{3}{quillen@pas.rochester.edu}
\altaffiltext{4}{blackman@pas.rochester.edu}

\begin{abstract}
We investigate the evolution of fossil cavities produced by extinct
YSO jets and wide angle outflows.  Fossil cavities are ellipsoidal or
cylindrical shells of swept-up of ambient (molecular cloud) material
moving at low velocities. The cavities form when the momentum in a YSO
jet or wide angle outflow decays in time allowing the bowshock or
swept-up shell to decelerate to velocities near the turbulent speed in
the cloud. It has been suggested in previous studies that cavities
provide efficient coupling between the jets/outflows and the cloud
and, as such, are the agents by which cloud turbulence is can be
re-energized. In this paper we carry forward a series of numerical
simulations of jets and outflows whose momentum flux decrease in
time. We compare simulations with decaying momentum fluxes to those
with constant flux.  We show that decaying flux models exhibit
deceleration of the outflow head and backfilling via expansion off of
the cavity walls. They also have lower density contrast, are longer
lived and wider than their continuously driven counterparts. The
simulations recover the basic properties of observed fossil cavities.
In addition, we provide synthetic observations in terms of P-V
diagrams which demonstrate that fossil cavities form both jets and
wide angle outflows are characterized by linear ``Hubble-law''
expansions patterns superimposed on ``spur'' patterns indicative of
the head of a bow shock.

\end{abstract}


\section{Introduction}
Energetic outflows from young stellar objects exert a strong effect on
their parent molecular clouds.  Consideration of the combined energy
budget for many outflows in dense star forming regions compared with
the energy of the parent molecular cloud's turbulent motions support
notions of feedback by showing an approximate balance between outflow
energy input and turbulent support
\citep{Bally,ballyreview,Knee,Matzner 2002,Warin}.  Outflow momentum
must be subsumed and isotropized into the cloud if young stellar
object outflows are to drive turbulence in molecular clouds.  The
explicit mechanisms by which this occurs has yet to be demonstrated.

\cite{Quillen} have recently explored the relation between turbulent
cloud motions and observed outflow activity in NGC 1333.  Their
results showed that the location of active outflows did not affect the
degree of turbulence in the cloud.  In addition recent numerical
studies \citep{Cunningham} have shown that direct collisions of active
outflows are not effective at transferring jet/outflow momentum to
cloud material (by active we mean that the momentum flux of the
outflow remains roughly constant through out the interaction). While
active outflows did not seem to energize turbulent motions
\cite{Quillen} were able to identify many sub-parsec size slowly
expanding cavities in 13CO observations of NGC 1333 which they
identified as dense wind-swept shells of gas that expand into the
cloud after the driving source of the outflow has expired. They found
that the mechanical energy required to open these cavities represents
a significant fraction of that required to drive the cloud
turbulence. The authors argued that these cavities were relics of
previously active molecular outflows.  Because cavities have a longer
lifetime than those opened by active and younger outflows, they would
be more numerous than higher velocity, younger outflows. These
cavities could provide the coupling between outflows and turbulence in
molecular clouds. Thus one means of re-energizing turbulence in clouds
may be through these ``fossil outflow'' cavities which are disrupted
after they have slowed to speeds comparable to the turbulent velocity
of the ambient cloud.

Our simulations focus on outflows whose energy and momentum flux
decreases with time. Such behavior is expected from observations which
show an exponential decrease in outflow momentum flux with age of
central source \citep{Bontemps}.  We note that most simulations of
molecular outflows have focused on continuous momentum injection
models (e.g., \citealt{suttner97,smith97,micono00,lee01,rosen03,
downes03,raga04,keegan05}). As an object ages, the bowshock driven by
a previously active outflow continues to expand into the molecular
cloud evolving into what \cite{Quillen} refer to this as a ``fossil
cavity''.  Theoretical models for expanding spherical cavities {\it
i.e.} wind blown bubbles, have been applied to cavities opened by
outflows in a few cases. \citet{koo92a} applied their wind blown
bubble model to the HH 7-11 region, located in the NGC 1333 cloud. In
the context of a slow wind, \citet{koo92a} suggested that cavities
opened in molecular clouds by outflows were consistent with their
estimated mechanical luminosities.

The observational properties of molecular outflows driven by jets and
winds have been modeled by many authors. But, as noted above, the
emphasis has most often been on active outflows. For example,
\citet{raga04} recently modeled the limb-brightened cavity associated
with HH 46/47 \citep{noriega04} as a bow shock driven by a perfectly
collimated active jet. Kinematic patterns of the swept-up material is
often of particular interest in these studies as these are readily
observable and can distinguish between models outflow formation.
Linear increases of velocity with distance, so-called Hubble Law
patterns, have been identified in some but not all theoretical
studies.  The bow shock model of \citet{zhang97} and the numerical
simulations of \citet{smith97,downes99,pol04} in which outflows are
created by the entrainment of ambient gas by a single bow shock all
reproduce the Hubble velocity law. In these models, the Hubble
velocity law is partly due to the geometry of the bow shock. An
explanation is that the highest forward velocities are found in the
apex of the bow shock, which is also the point farther away from the
source. The forward velocities decrease toward the wings of the bow
shock, and the farthest away from the apex (or closer to the outflow
source), the slower will be the entrained gas (see figure 3 of
\citealt{masson}, figure 12 of \citealt{lada96} \& figure 4 of
\citealt{welch}).  In other models, however, Hubble laws do not appear
as a consequence of either analytical or numerical simulations. In the
studies of \citet{ostriker01,lee01} jets produce a characteristic
``spur'' pattern in the swept-up gas. It seems that some combination
of entrainment or unstable boundaries may be needed to create Hubble
laws in active outflows.

In this paper we carry out a numerical exploration of cavities that
are opened in molecular clouds by both jets and wide angle winds
(WAW). We compare the morphology and velocity structure of cavities
caused by continuously injected momentum with those of extinct
outflows.  Our first goal is to recover observable properties of
cavities. This is the first step in verifying the conjecture that
fossil cavities represent the means of transferring outflow energy and
momentum into cloud turbulence.  By confirming that we can recover the
morphology and kinematics of observed fossil shells we can then, in
future works, explore cavity-cavity collisions and the energizing of
turbulent motions.  We also use our simulations to identify physical
differences between continually driven and extinct outflow driven
structures and their impact on observables such as PV diagrams.  In
section II we present our method and model for the jet/WAW source.  In
section III we present our results. Section IV we present a discussion
of the results in light of outstanding issues in molecular outflows
and issues of protostellar turbulence.

\section{Computational Method and Initial Conditions}
\subsection{Method: Numerical Code}
The simulations presented here have been carried out in 2D cylindrical
axial symmetry (2.5D) using the AstroBEAR adaptive mesh refinement
(AMR) code. AMR allows high resolution to be achieved only in those
regions which require it due to the presence of steep gradients in
critical quantities such as gas density. The hydrodynamic version of
AstroBEAR has been well tested on variety of problems in 1, 2, 2.5D
\citep{pol,var,Cunningham} and 3D \citep{Lebedev 2004}.

For the results presented here, AstroBEAR was used to evolve Euler's
equations including the effects of time dependent radiative cooling
$H_2$ chemistry, $H$ and $He$ ionization.  The implementation of
radiative cooling, chemistry and ionization employed are discussed in
detail in \cite{Cunningham}.  A spatial and temporal second order
accurate MUSCL scheme using a Roe-average linearized Riemann solver
was used to integrate the source-free Euler equations.  The MUSCL
scheme employed achieves second order spatial accuracy by performing a
MINMOD interpolation of the primitive fields (density, velocity and
pressure) to grid interfaces.  The TVD-preserving time stepping method
of \cite{Shu 1988} is used to advance the solution.  The
micro-physical source terms are handled separately from the
hydrodynamic integration using an operator split approach.  The source
term is integrated using an implicit fourth-order Rosenbrock
integration scheme for stiff ODE's. We have made use of the Local
Oscillation Filter method of \cite{sutherland} using a viscosity
parameter $\alpha=0.075$ to eliminate numerical instabilities that can
occur near strongly radiative shock fronts.

\subsection{Model: Fossil Cavity Simulations}
We have performed simulations for the case where the outflow source
expires and creates a fossil cavity and also for the case of a
continuously driven source.  The simulations were carried forward for
two outflow sources; the case of a collimated jet and the case of a
wide angle wind that was injected into the domain along the surface of
a sphere where the density of the driving source varies with polar
angle $\theta$ as $1/\max(\sin^2(\theta),0.0025)$.  For numerical
palatability, we have constructed this function to have a limiting
value of $400$ near the pole.  This model has been shown to be a good
approximation of X-winds \cite{shu95}.  Furthermore, \cite{matzner99}
have argued that this model is characteristic of any radial
hydromagnetic wind at distances far from the driving source.
Hydrodynamic winds of this form, we have found, produce relatively
collimated outflows in both uniform and toroidal density environments.
The parameters common to each of the four simulations prescribed here
are presented in table \ref{t1}.  For the cases where the driving
source expires, the density of the wind is decreased as:
\[
\rho(t) = \left\{
\begin{array}{ll}
\rho_0 \exp\left[-\left(t/\tau_{decay}\right)^2\right],&\mbox{ if } t < t_{off} \\
\rho_0 \exp\left[-\left(t_{off}/\tau_{decay}\right)^2\right],&\mbox{ otherwise }
\end{array}\right.
\]
and the velocity of the driving wind is reduced to zero as:
\[
v(t)=\left\{
\begin{array}{ll}
v_o,&\mbox{ if } t \le 0.95~t_{off} \\
v_o \left[1-\frac{t-0.95~t_{off}}{0.1~t_{off}} \right],&\mbox{ if } 0.95~t_{off} < t < 1.05~t_{off} \\
0,&\mbox{ otherwise. }
\end{array}\right.
\]
The jet outflow conditions were imposed from within the boundary zones
inside of a disk of radius $r_j$, centered on the origin. Similarly, the wide
angle outflow was imposed within a sphere of radius $r_j$ centered at
the origin.  Reflection boundary conditions were used outside of the
outflow source along $z=0$ and $r=0$ and extrapolation boundary
conditions were imposed elsewhere.
%\clearpage
\begin{table}[!h] \caption{Simulation Parameters. \label{t1}}
 \begin{tabular}{l l}
 \tableline
 Domain Size, $(Z \times R)$,& $1~\textrm{pc} \times 0.1667~\textrm{pc}$ (jet) \\
                             & $1~\textrm{pc} \times 0.2083~\textrm{pc}$ (WAW) \\
 Jet Radius, $r_j$ & $0.0208~\textrm{pc} = 4730~\textrm{AU}$ \\
 Computational cells per $r_j$ & $64$ \\
 Resolution & $1\times10^{15}~\textrm{cm}$ \\
 Outflow Rate, $\dot{M}$ & $1\times10^{-6}~\textrm{M}_{\sun}~\textrm{yr}^{-1}$ \\
 Outflow Velocity, $v_0$ & $100~\textrm{km s}^{-1}$ \\
 Outflow Temperature & $10^4~\textrm{K}$ \\
 Outflow Gas Composition & 75\% $HII$, 25\% $HeIII$ by mass \\
 Ambient Density $\rho_a$ & $1000~\textrm{cm}^{-3}$ \\
 Ambient Temperature & $50~\textrm{K}$ \\
 Ambient Gas Composition & 75\% $H_2$, 25\% $HeI$ by mass \\
 Decaying Wind Shut-off Time, $t_{off}$ & $9.5~\textrm{kyr}$ (jet) \& 
                       $6.25~\textrm{kyr}$ (wide angle wind) \\
 Decaying Wind Shut-off Decay Time, $\tau_{decay}$ & $4.75~\textrm{kyr}$ (jet) \& 
                       $3.125~\textrm{kyr}$ (wide angle wind) \\
 \tableline
 \end{tabular}
\end{table}
%\clearpage
Observations by \cite{Bontemps} reveal that the momentum flux supplied
to protostellar outflows decreases by two orders of magnitude as the
progenitor evolves from and embedded class 0 source to an evolved
class I source in a manner that is in rough proportion to the
circumstellar envelope mass.  Our model for the time dependence of the
outflow source is consistent with this observed decrease.  The fossil
outflow cavity models presented here reduce the density of the driving
wind to $2\%$ of its initial value at $t=t_{off}$ while maintaining
the wind velocity near the escape speed.  The driving wind is shut off
($v_w=0$) for $t>t_{off}$ to model the effect of a completely extinct
outflow source.  While the driving winds in our cavity models cease
approximately $10\times$ earlier than the typical age of class I
sources, the total momentum flux injected in our fossil cavity modes
$\chi=0.441 \dot{M} v_0 t_{off} \sim
1~\textrm{M}_{\sun}~\textrm{yr}^{-1}$ is characteristic of the net
momentum flux inferred from observation of low mass sources
\citep{Knee}. The combination of detailed microphysics and long
simulation time make the simulations computationally demanding.  We
acknowledge that the radius of the outflow source used in our
numerical models of $4300~\textrm{AU}$ is $\sim 40\times$ that of
typical collimated outflows.  The outflows driven here are therefore
very light compared to more collimated models with comparable mass
loss rates.  We chose to accept this trade-off in order to enable the
simulation of large scale outflow cavities over long timescales while
resolving the embedded outflow source.  Note also that we use only a
hydrodynamic version of the code as the magnetic field strengths
inferred in jets at these size scales are unlikely to alter the
dynamics of the global flow patterns \citep{crutcher}.

\section{Results}
\subsection{Outflow \& Cavity Morphology}
\begin{figure}[!h]
\begin{center}  
\includegraphics[angle=-90,clip=true,width=0.667\textwidth]{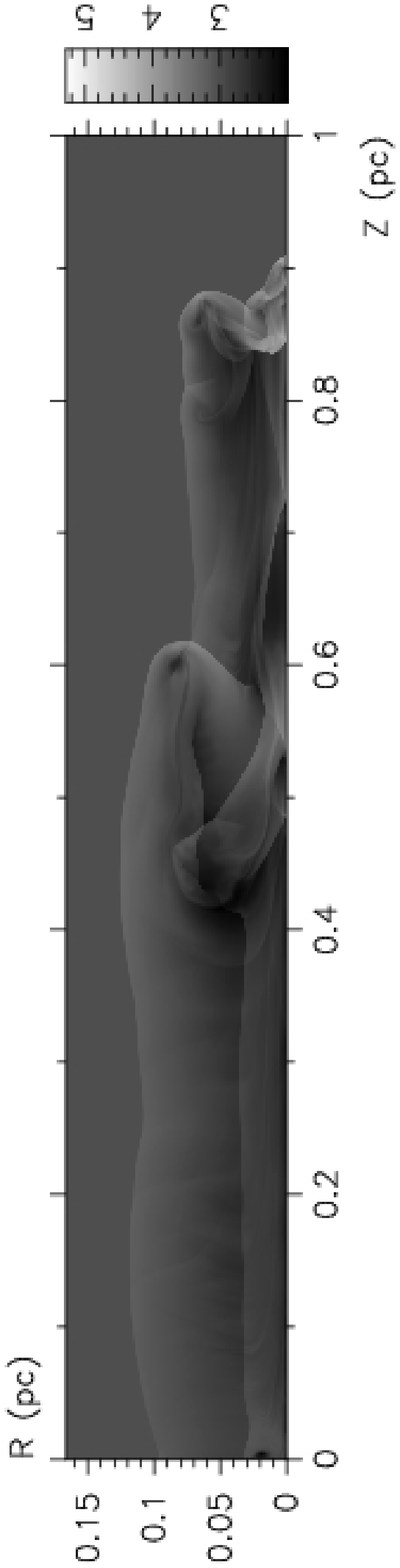}\\
\includegraphics[angle=-90,clip=true,width=0.667\textwidth]{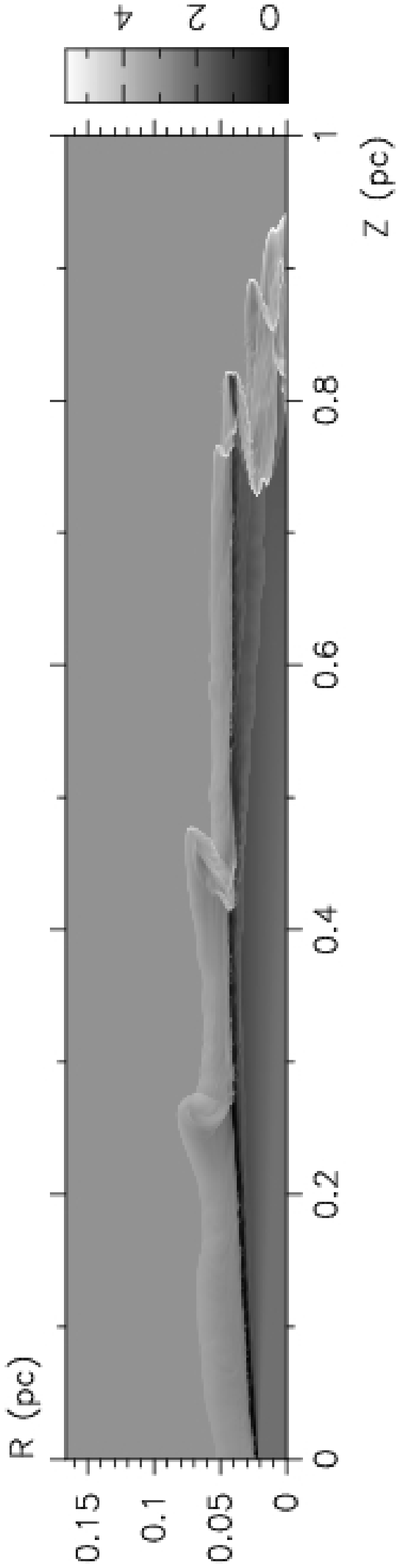}
\end{center}
\caption{Grayscale density images of a jet driven outflow with the
driving source was turned off at $9.5~\textrm{kyr}$ shown at
$100~\textrm{kyr}$ (top) and continuously driven shown at
$34~\textrm{kyr}$ (bottom).  The fossil cavity (top) with higher
density contrast than the continuously driven outflow (bottom).
\label{f1}}
\end{figure}

\begin{figure}[!h]
\begin{center}
\includegraphics[angle=-90,clip=true,width=0.667\textwidth]{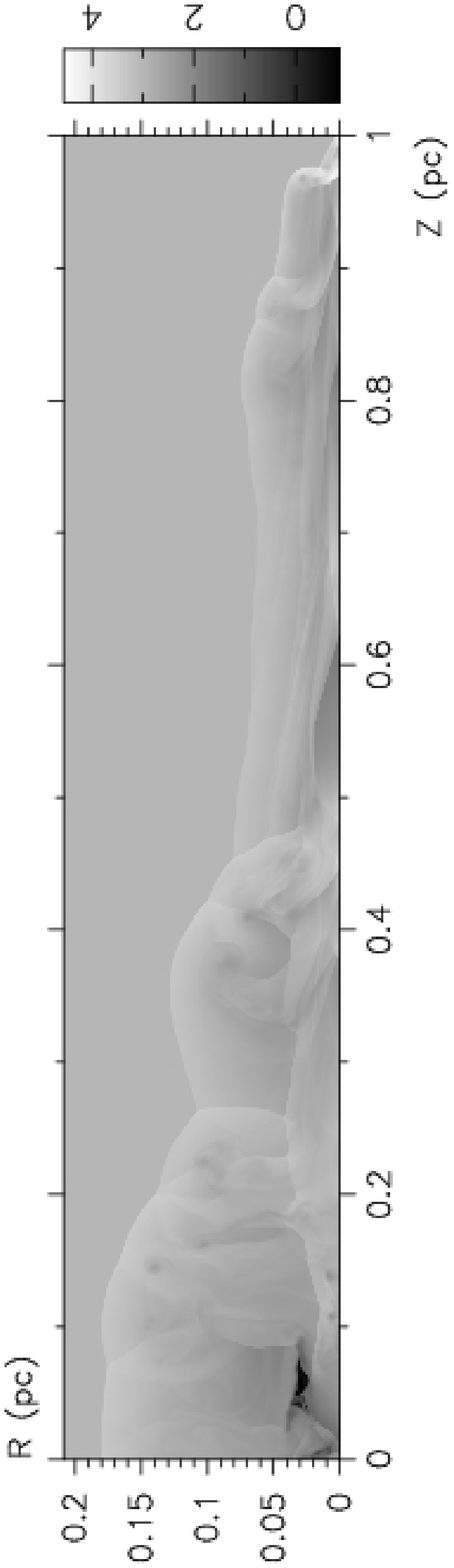}\\
\includegraphics[angle=-90,clip=true,width=0.667\textwidth]{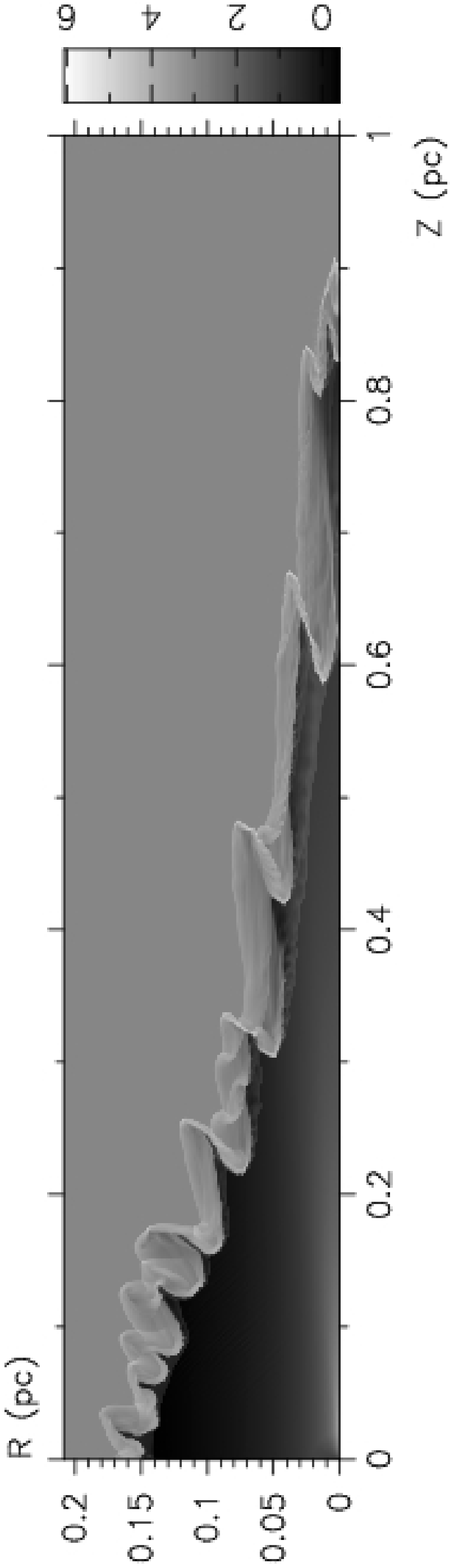}
\end{center}
\caption{Grayscale density images of a wide angle wind driven outflow
with the driving source was turned off at $6.25~\textrm{kyr}$ shown at
$89~\textrm{kyr}$ (top) and continuously driven shown at
$33~\textrm{kyr}$ (bottom).
\label{f2}}
\end{figure}

We now present the results of the simulation suite. Figure \ref{f1}
(top) shows the cylindrically symmetric simulation of a
$10^5~\textrm{yr}$ old fossil outflow cavity that has been opened by a
collimated jet.  The density of the driving source was decreased until
the driving source was turned off at $t_{off}=9.5~\textrm{kyr}$.  For
comparison the simulation where the inflow rate and velocity remain
continuous is shown in figure \ref{f1} (bottom).  Figure \ref{f2}
shows the wide-angle simulation set with a decaying wind (top) and
continuous outflow (bottom).  In the wide angle wind case, the driving
source was turned of at $t_{off}=6.25~\textrm{kyr}$.  As noted
earlier, the prescription for the wide angle wind used here produces a
relatively collimated outflow structure owing to the divergence in
wind density at the poles.  The irregular bow shock shape of the jet
driven models is due to vortex shedding from the condensed gas at the
head of the jet.  We note that the condensed layer of strongly cooling
wind-swept gas at the head of the jet and along the entire
circumference of the wide angle wind continuously driven cases
(figures \ref{f1} \& \ref{f2}, bottom) are subject to non-linear
bending mode instabilities \citep{vishniac}.  We have performed these
simulations at $2\times$ lower resolution to assess the convergence of
these results.  The lower resolution results also feature unstable
thin shells and vortex shedding but the details of the small scale
features associated with these phenomena are not converged.  However,
the overall shape and dynamics of the flows that are pertinent to the
results presented in this paper are unchanged at lower resolution.

\begin{figure}[!h]
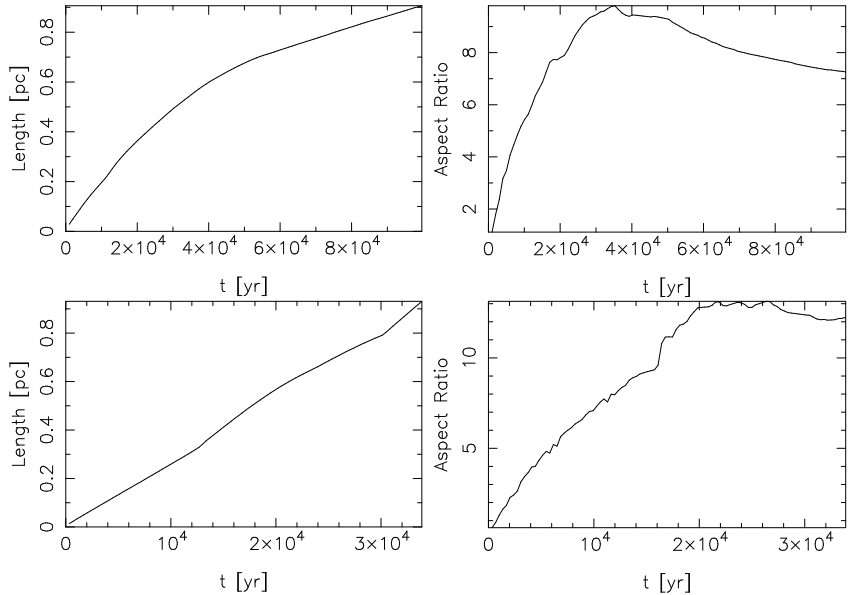

\begin{center}
  \includegraphics[angle=-90,clip=true,width=0.333\textwidth]{f3a.ps}
  \includegraphics[angle=-90,clip=true,width=0.333\textwidth]{f3b.ps}\\
  \includegraphics[angle=-90,clip=true,width=0.333\textwidth]{f3c.ps}
  \includegraphics[angle=-90,clip=true,width=0.333\textwidth]{f3d.ps}
\end{center}
\caption{ Shape of the jet driven fossil cavity (top) and continuously
driven outflow (bottom).  The cessation of the driving source is
apparent as a change in slope at $t \sim 10^4~\textrm{yr}$ on the top
left.  While the aspect ratio approaches a constant value for the
fossil cavity (top right), it continues to increase for the
continuously driven jet (bottom right).
\label{f3}}
\end{figure}

\begin{figure}[!h]
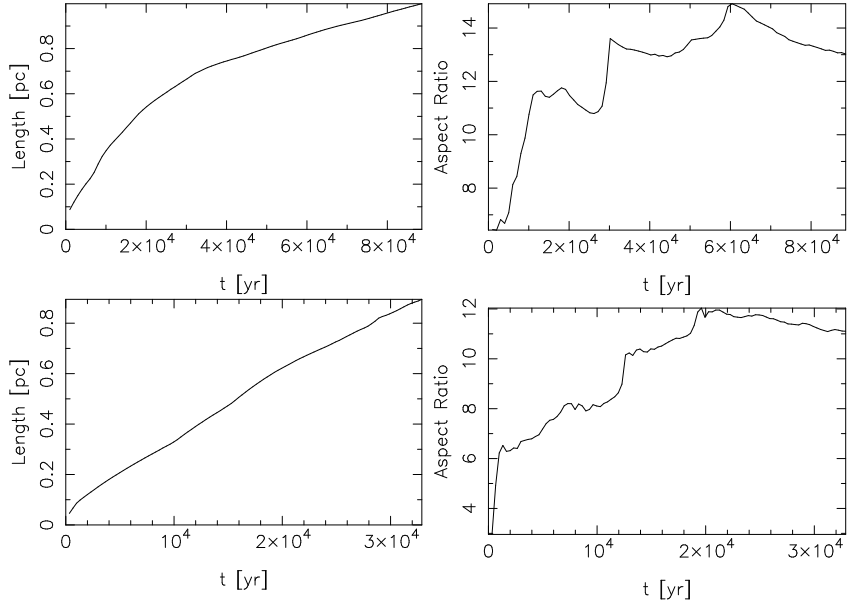

\begin{center}
  \includegraphics[angle=-90,clip=true,width=0.333\textwidth]{f4a.ps}
  \includegraphics[angle=-90,clip=true,width=0.333\textwidth]{f4b.ps}\\
  \includegraphics[angle=-90,clip=true,width=0.333\textwidth]{f4c.ps}
  \includegraphics[angle=-90,clip=true,width=0.333\textwidth]{f4d.ps}
\end{center}
\caption{ Shape of the wide angle wind driven fossil cavity (top) and
continuously driven outflow (bottom). The cessation of the driving
source is apparent as a change in slope at $t \sim 10^4~\textrm{yr}$
on the top left. While the aspect ratio is nearly constant for the
fossil cavity (top right), it continues to increase for the
continuously driven wind (bottom right).
\label{f4}}
\end{figure}

The fossil cavities opened after the driving jet or wind expires have
similar spatial extents, $L \sim 1~\textrm{pc}$ and age, $t_f \sim
10^5~\textrm{yr}$.  The continuously driven outflows are shown at
times of comparable spatial extent to their fossil counterparts
($1~\textrm{pc}$) but are considerably younger at $\sim 1/3$ the
age. The difference is attributable to the rapid deceleration of the
leading edge of the fossil outflows after the driving source has
expired.
% Figure 1
%\clearpage
% Figure 2
%\clearpage
The morphological signatures in terms of aspect ratio of fossil
outflow cavities differ from their continuously driven counterparts.
The propagation of the head of the outflow quickly decelerated to $<
10~\textrm{km s}^{-1}$ in less than $10~\textrm{kyr}$ after the
driving source expired (figures \ref{f3} \& \ref{f4}, left).  We
define the outflow aspect ratio as the ratio of the outflow length to
the width of the outflow at its half-length.  The outflow aspect ratio
is plotted as a function of time for the jet and wide angle wind cases
in figures \ref{f3} \& \ref{f4}, right, respectively.  Fossil outflow
cavities evolve to be $\sim 50\%$ wider, relative to their length than
their continuously driven counterparts.  After the driving source
expires, the rate of increase of the aspect ratio diminishes as the
head of the outflow decelerates more rapidly than the lateral
edges. This can be seen in figure \ref{f4} (top) for the jet driven
cavity. The rapid deceleration of the outflow after the driving source
has turned off at $9.5~\textrm{kyr}$ is apparent in the plot of outflow
length shown in figure \ref{f3}, top left.  The length to width ratio
plateaus to $8$ as the head of the cavity continues to decelerate.
For the continuously driven jet case (figure \ref{f3}, bottom) the
propagation of the head of the outflow continues at a roughly constant
rate.  The length to width radio plateaus to $12$ at late time due to
an event of vortex shedding at the head of the outflow.  If the
simulation were to continue, we expect that the the length to width
ratio would resume the constant rate of increase as present at earlier
times, before $t \sim 1.4\times10^4~\textrm{yr}$.

The propagation of the head of the wide angle wind driven outflow
(figure \ref{f4}) also undergoes rapid deceleration after the driving
wind has been turned off at $6.25~\textrm{kyr}$.  The length to width
ratio of the outflow cavity plateaus to $14$ after the driving source
has expired.  In this case the length to width ratio to undergo a
brief periods of reduction after $20~\textrm{kyr}$ owing to the
irregularly shape of the bow shock.  This decrease is due to the
geometry of the wide angle wind driven cavity.  After the driving
source expires the outflow cavity, initially wider at the base than at
the head, expands to a more cylindrical geometry as the highly
supersonic head of the cavity decelerates more rapidly than the radial
wall.
% Figure 3
%\clearpage
% Figure 4
%\clearpage
\subsection{Fossil Cavity Backfilling}
The interior walls of an outflow cavity are supported, in the case of
a wide angle wind by the ram pressure of outflowing material
\citep{delamarter}.  In the case of a jet source the interior cavity
wall is supported at the head by the ram pressure of the jet and along
the radial edges by shock heated gas in the ``cocoon'' region between
the jet beam and bow shock \citep{blondin}.  The jet models presented
here feature a tenuous, narrow cocoon region between the radial edge
of the jet beam and the dense shell of wind-swept ambient gas due to
strong radiative cooling (figure \ref{f1}, bottom).  The radiative
cooling acts to remove the pressure support of the radial edges of the
cavity.  The cocoon pressure plays dynamically minor in these models.

\begin{figure}[!h]
\begin{center}
  \begin{tabular}{l}
  \hline
  \includegraphics[angle=-90,clip=true,width=0.6\textwidth]{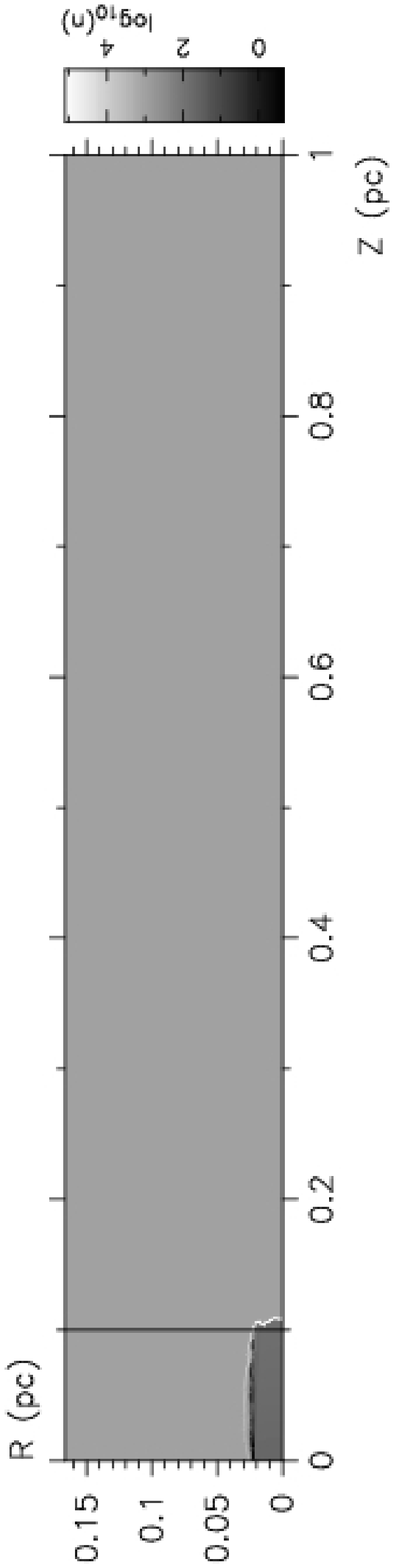} \\
  \includegraphics[angle=-90,clip=true,width=0.3\textwidth]{f5b.ps}
  \includegraphics[angle=-90,clip=true,width=0.3\textwidth]{f5c.ps} \\
  \hline
  \includegraphics[angle=-90,clip=true,width=0.6\textwidth]{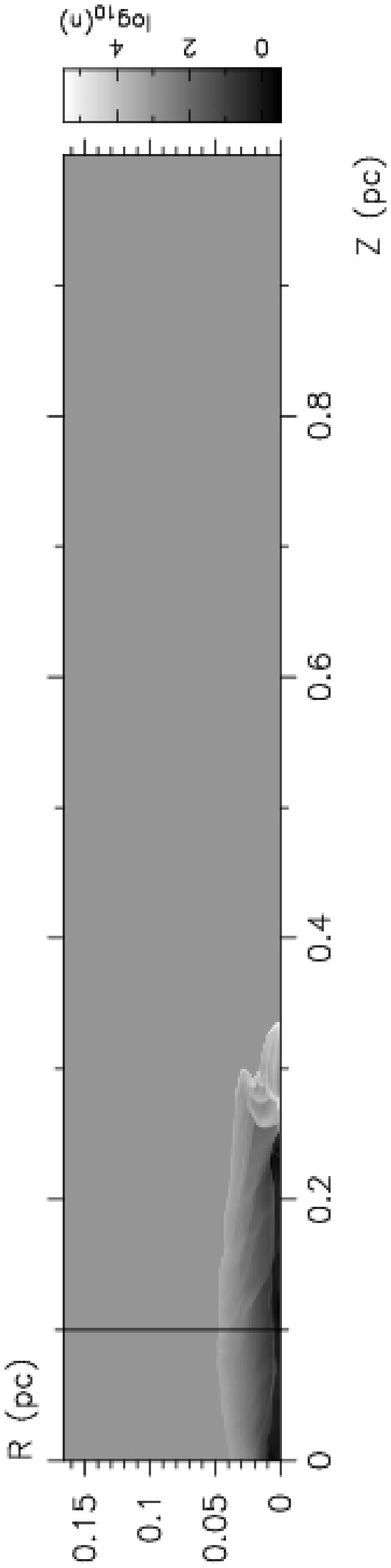} \\
  \includegraphics[angle=-90,clip=true,width=0.3\textwidth]{f5e.ps}
  \includegraphics[angle=-90,clip=true,width=0.3\textwidth]{f5f.ps} \\
  \hline
  \includegraphics[angle=-90,clip=true,width=0.6\textwidth]{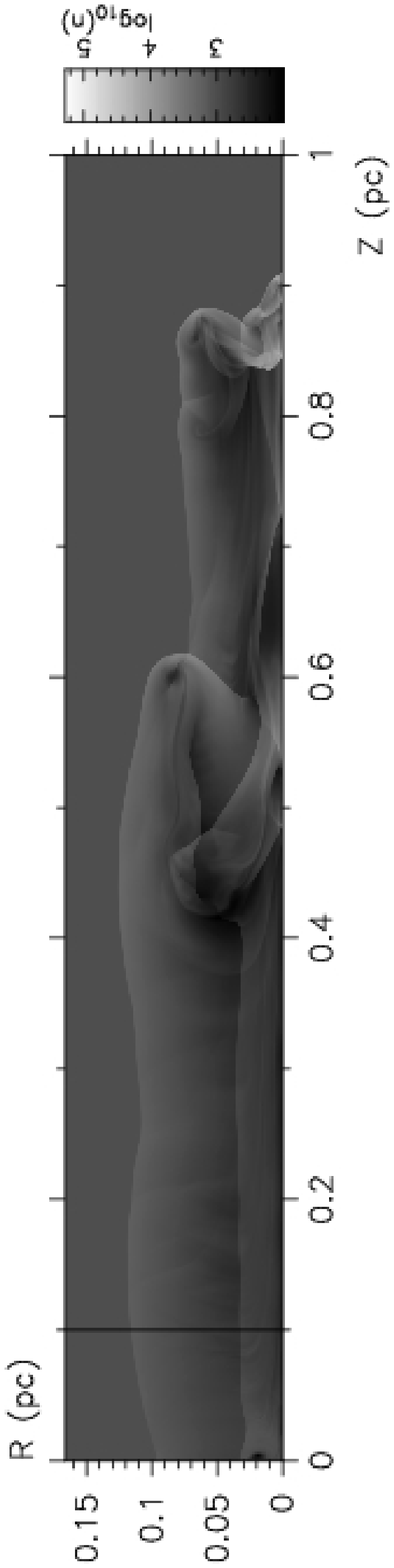} \\
  \includegraphics[angle=-90,clip=true,width=0.3\textwidth]{f5h.ps}
  \includegraphics[angle=-90,clip=true,width=0.3\textwidth]{f5i.ps} \\
  \hline
  \end{tabular}
\caption{Three panels show crosscuts of the fossil collimated jet
cavity simulation at time $t=5~\textrm{kyr}$ (top),
$t=18~\textrm{kyr}$ (center) and $t=100~\textrm{kyr}$ (bottom).  Each
panel is composed of density grayscale images (top) showing the
position of the crosscuts of density (lower left), $v_r$ (solid line),
$v_z$ (dotted line) and $\pm$ sound speed (dashed line, lower right).
This figure illustrates the dynamic infall of the cavity wall after
the driving source expires.
\label{f5}}
\end{center}
\end{figure}

\begin{figure}[!h]
\begin{center}
  \begin{tabular}{l}
  \hline
  \includegraphics[angle=-90,clip=true,width=0.6\textwidth]{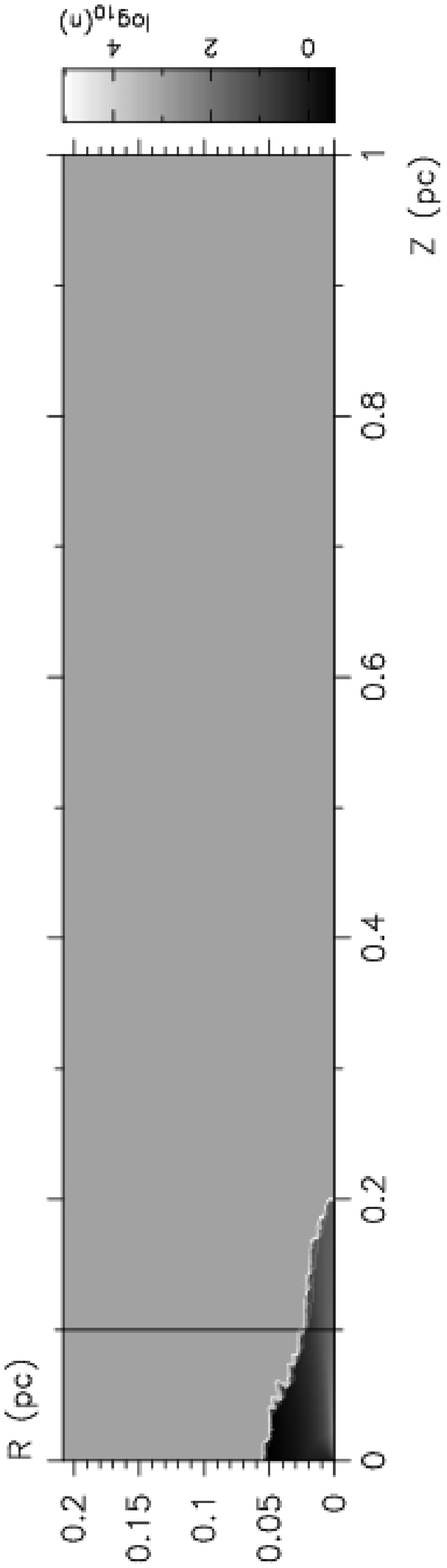} \\
  \includegraphics[angle=-90,clip=true,width=0.3\textwidth]{f6b.ps}
  \includegraphics[angle=-90,clip=true,width=0.3\textwidth]{f6c.ps} \\
  \hline
  \includegraphics[angle=-90,clip=true,width=0.6\textwidth]{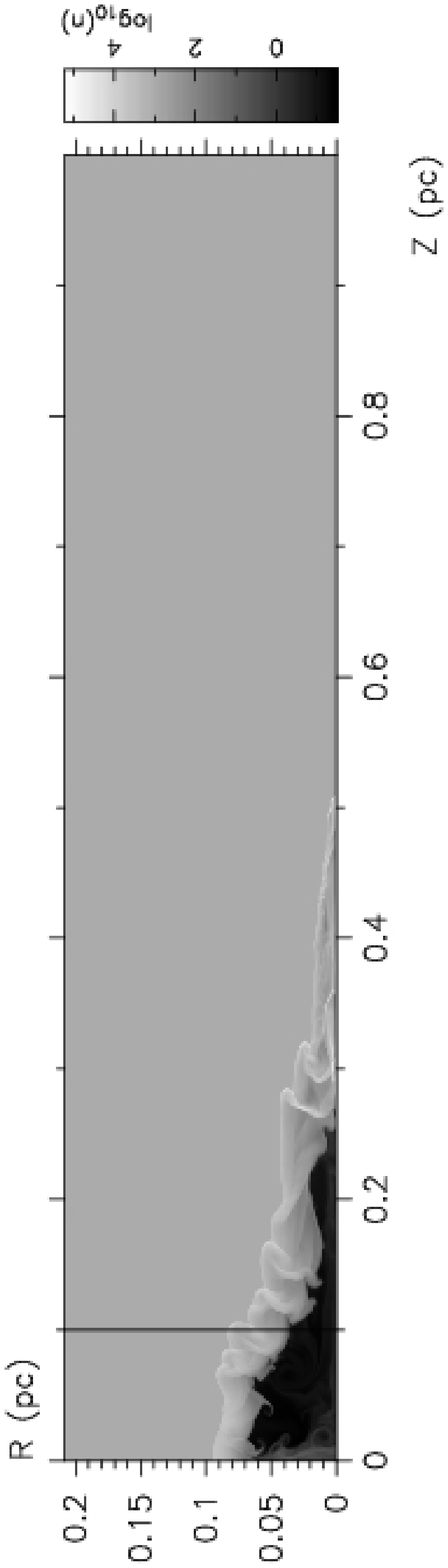} \\
  \includegraphics[angle=-90,clip=true,width=0.3\textwidth]{f6e.ps}
  \includegraphics[angle=-90,clip=true,width=0.3\textwidth]{f6f.ps} \\
  \hline
  \includegraphics[angle=-90,clip=true,width=0.6\textwidth]{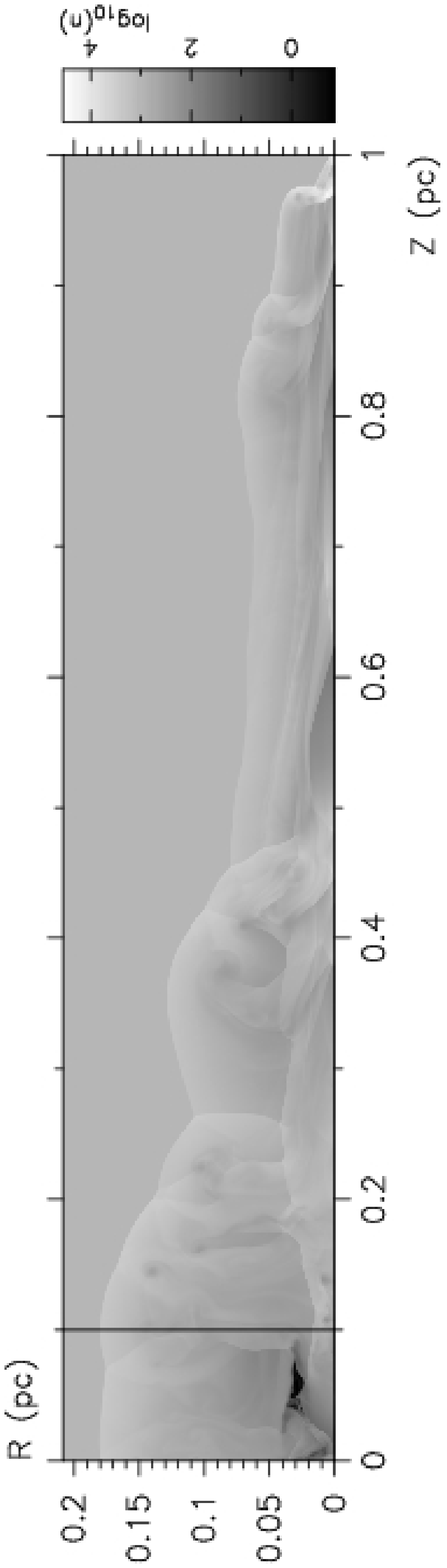} \\
  \includegraphics[angle=-90,clip=true,width=0.3\textwidth]{f6h.ps}
  \includegraphics[angle=-90,clip=true,width=0.3\textwidth]{f6i.ps} \\
  \hline
  \end{tabular}
\end{center}
\caption{Three panels show crosscuts of the fossil wide angle wind
cavity simulation at time $t=5~\textrm{kyr}$ (top),
$t=18~\textrm{kyr}$ (center) and $t=100~\textrm{kyr}$ (bottom).  Each
panel is composed of density grayscale images (top) showing the
position of the crosscuts of density (lower left), $v_r$ (solid
line), $v_z$ (dotted line) and $\pm$ sound speed
(dashed line, lower right).
\label{f6}}
\end{figure}

When the driving source shuts off the walls of the cavity and will
flow back towards the source through an expansion wave.  The backfill
process can be seen in the fossil cavity as a reduced contrast in the
cavity compared with the continuous cases.  Figures \ref{f5} \&
\ref{f6} plot crosscuts taken transverse to the direction of outflow
direction showing $\rho$, $v_r$ (solid line), $v_z$ (dotted line), and
$\pm c_s$ (dashed line) for the jet and wide angle driven fossil
cavities respectively.  The backfill process is driven by a triplet of
waves.  1) An interior or ``wind'' shock delineating the bounds of the
cavity evacuated after the expiration of the wind source, 2) a contact
discontinuity separating wind gas from wind-swept ambient gas, and 3)
an expansion wave that drives the backfill of the dense shell of
wind-swept gas toward the interior of the cavity.  The speed of the
inward collapse of wind-swept gas is expected to be approximately that
that of an expanding rarefaction, $v_{exp} = \frac{2}{\gamma-1} c_s$.
The low-latitude edges of the cavity retain some forward or axial
momentum. Therefore the walls of the cavity will diagonally forward
with negative $v_r$ and positive $v_z$.

In the top crosscut panel of figure \ref{f5} at $5~\textrm{kyr}$, the
outflow source is still driving the jet.  The density cross cut
reveals a $\sim 10 \times$ density contrast (the ratio of the ambient
density to that inside the cavity).  As expected the velocity plots
reveal that the flow inside the jet beam diving the outflow is
hypersonic and forwardly directed with subsonic radial motion in the
cocoon between the driving jet and bow shock.  At $18~\textrm{kyr}$
(center panel), the rarefaction that ensues due to the expiration of
the driving wind has left an evacuated cavity with a large ($\sim 1000
\times$) density contrast in the region previously occupied by the
driving wind.  At this time the dense shell of wind swept gas just
behind bow shock has begun backfilling into the cavity at $v \sim c_s
\sim 1~\textrm{km s}^{-1}$.  The interior shock that precedes the
infall expansion wave has not yet reached the symmetry axis of the
cavity.  In the lower panel taken at $100~\textrm{kyr}$ the interior
shock has reflected off the symmetry axis.  The cavity walls have
completely backfilled into the cavity leaving a density contrast of
$\sim 2$.  At this evolved stage, the motion inside the cavity is
primarily radial infall.  The density contrast of the cavity will
become further reduced as the infall of the cavity walls continue at
later time.

Figure \ref{f6} shows crosscuts for the wide angle wind simulation.
In the top panel, taken at $5~\textrm{kyr}$, the wide angle wind
outflow source is still active.  At the polar latitudes intercepted by
the crosscut, most of the velocity is forwardly directed.  The densest
outflowing gas appears in the swept-up shell and the in the ``jet''
associated with the flow along the pole (high latitudes). The maximum
$\sim 100 \times$ density contrast occurs inside the cavity at lower
latitude.  At $18~\textrm{kyr}$ (center panel), the rarefaction that
ensues due to the expiration of the driving wind has left an evacuated
cavity with a density contrast of $\sim 1000 \times$.  The outer shell
of wind-swept gas backfills at $v \sim 1~\textrm{km s}^{-1}$.  In the
lower panel at $89~\textrm{kyr}$ the interior shock has reflected off
the symmetry axis.  At this time, the maximum density contrast inside
the cavity has dropped to $< 5$.
% Figure 5
\clearpage
% Figure 6
\clearpage
\setlength{\voffset}{-15mm}
\clearpage
\setlength{\voffset}{0mm}
\subsection{Cavity Deceleration and Protostellar Turbulence}
We seek a simple analytic estimate relating the age and momentum
carried by fossil cavities to their directly observable properties
(size, mass, expansion rate).  Consideration of a spherically
symmetric impulse carrying total momentum $\chi$ imparted into a thin
shell of wind swept gas that propagates into a uniform environment of
density $\rho_a$ leads to a momentum conserving self-similar model for
the the observed outflow cavity.  The volume, $V_{shell}$, swept by
the shell is given as
\[
V_{shell}=\frac{4\pi}{3}(v(t)t)^3
\]
where $v(t)$ is the velocity of the expanding shell.  Momentum
conservation yields
\[
V_{shell}=\frac{4\pi}{3}\left(\frac{\chi}{M} t\right)^3,
\]
where $M=\rho_a V$is the mass entrained into the wind swept shell.
Substituting for the mass swept into the shell we have
\[
V^{4/3}=\left(\frac{4\pi}{3}\right)^{1/3}\frac{\chi}{\rho_a} t.
\]

This scaling relation provides a simple model for momentum-driven,
self-similar, spherical flows.  \cite{Quillen} employ this scaling
relation to infer that the ensemble of fossil outflow cavities they
observed in NGC 1333 contain enough momentum to maintain turbulence in
the cloud.  This relation applies only to extinct outflows with
$t>>t_{off}$, due to the explicit assumption of an impulsive driving
wind built into the scaling relation.  While protostars do not eject
outflow momentum instantaneously, this is a reasonable approximation
for long-extinct outflows.  We emphasize that due to the strong
radiative energy loss characteristic of YSO environments, thermal
pressure plays a minor role in driving their dynamics.  Therefore such
flows satisfy the momentum-driven constraint.  However, YSO outflows,
in general, are not spherical and the backfill of the fossil cavity
cannot be described by such a self-similar motion.  We therefore
express the volume $V$ subtended by the bow shock of a YSO outflow
cavity more generally as
\[
V^{4/3}=f(t) \left(\frac{4\pi}{3}\right)^{1/3}\frac{\chi}{\rho_a} t
\]
where $f(t)$ is an arbitrary function of time that accounts for the
deviation of the actual cavity size from that predicted by the
approximate scaling relation.

\begin{figure}[!h]
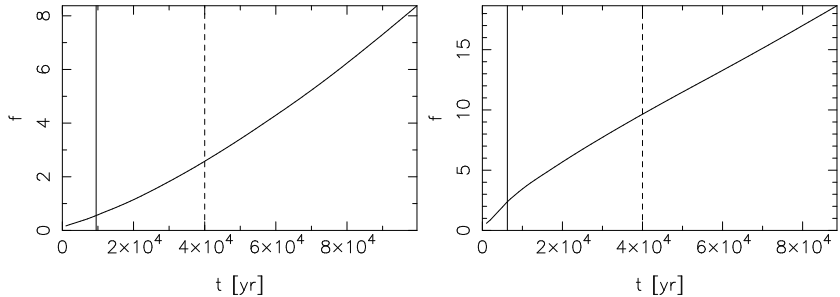

\begin{center}
  \includegraphics[angle=-90,clip=true,width=0.33\textwidth]{f7a.ps}
  \includegraphics[angle=-90,clip=true,width=0.33\textwidth]{f7b.ps}
\end{center}
\caption{ The correction factor $f(t)$ due to the deviation of the
size of the jet (left) and wide angle wind (right) driven cavities.  A
momentum conserving self-similar solution would have $f(t)=1$ for
$t>t_{off}$ (solid vertical line).  The radial expansion of the
cavities slows to $< 1~\textrm{km s}^{-1}$ by $40~\textrm{kyr}$
(dashed vertical line).
\label{f7}}
\end{figure}

We extract $f(t)$ from the simulations by measuring the volume
enclosed by the bow shock delineating the fossil cavity at 100 equally
spaced intervals in time through out the simulation.  The ambient
density and net momentum injection used in the fossil cavity models
presented here yield $\rho_a / \chi = 5.77 \times
10^7~\textrm{pc}^{-4}~\textrm{yr}$ for the jet model and $\rho_a /
\chi = 8.76 \times 10^7~\textrm{pc}^{-4}~\textrm{yr}$ for the WAW
model.  We have taken $\chi$ as the net momentum imparted to the
cavity throughout the simulation independent of time.  Therefore
$f(t)$ is the factor by which the impulsive, momentum-driven scaling
relation misinterprets the cavity size-age relationship of the
simulated fossil cavities.  Figure \ref{f7} shows 
\[
f(t)=\left(\frac{3}{4\pi}\right)^{1/3} \frac{V^{4/3} \rho_a}{\chi t}
\]
 for the jet driven (left) and WAW driven (right) cases.  The solid
vertical line in the plots is at $t=t_{off} = 9.5,~6.25~\textrm{kyr}$
for the jet and wide angle wind cases respectively and the dotted
vertical line in the plots at $t=40~\textrm{kyr}$ denotes the time
where the radial edge of the fossil cavities have decelerated to
$<1~\textrm{km s}^{-1}$. First we note that $f(t)\sim 1$ throughout
most of the cavity lifetime.  The faster than predicted cavity
expansion is likely due to a combination of factors not accounted for
in the simple scaling law. First is the effect of the geometry of the
flow.  The scaling law is essentially a 1D approximation while the
simulations show complex 2D motions both at the head of the jet and,
in the case of the wide angle wind, along the walls of the cavity
where wind material is redirected by an oblique shock. The collapse of
the interior walls into the cavity also changes the global dynamics.
This effect violates the assumption that a thin, wind swept shell of
gas will delineate the cavity perimeter.  The ambient gas overrun by
the cavity will not, in general, be accelerated to travel with the
outer outflow shell, particularly in the case of a jet-driven cavity.
Some of the gas entrained into a cavity may be accelerated to a speed
that is only some fraction of that of the cavity walls.  The scaling
law therefore under estimates cavity volume.  The most significant
observational consequence is that this model may underestimate the net
momentum of observed cavities inferred from their volume.  In spite of
these effects the scaling law retains reasonable predictive capacity
for interpreting decelerating cavities.  Because the geometry of the
driving wind of an observed cavity is often unknown, the scaling
relation provides a reasonable rough approximation for interpreting
observational data, particularly when the exact geometry of the cavity
is obfuscated.
% Figure 7
%\clearpage
%\clearpage
Figures \ref{f5} \& \ref{f6} show that the expansion of the radial
edges of the simulated cavities decelerated to considerably less than
the turbulent speed typical of star forming environments of a few
$\textrm{km s}^{-1}$ \citep{vaz} by $2 \times 10^5~\textrm{yr}$.
Therefore, the cavities can continue to expand for no more than a few
more turbulent crossing times.  By the time several turbulent crossing
times have elapsed, the cavity will have entrained enough of the
turbulent ambient gas to be subsumed into and become indistinguishable
form the turbulent motions of the cloud. Its momentum will act to feed
the turbulence that will ultimately destroy it.  \cite{Quillen} reach
the same conclusion for cavities observed in NGC 1333 using the
scaling relation that is in rough agreement with our simulation
models.  This picture of outflow driven turbulence is also supported
by recent 3D numerical MHD simulations of \cite{li06}.  They model
protostellar outflow ejection from dense condensations that form the
evolution of an initially Jeans-unstable, rotating, magnetized cloud.
The initially seeded ``interstellar turbulence'' decays rapidly,
allowing the formation of gravitationally bound condensations.  They
find that ``protostellar turbulence'' replenishes turbulent motion and
inhibits further collapse in the cloud by imposing that 10\% of each
core mass is ejected at speeds characteristic of protostellar
outflows.  We note in that study the ``fossil'' nature of the driving
outflows was not addressed.

The velocity of the wind in the simulations was reduced from
$100~\textrm{km s}^{-1}$ to zero rather abruptly.  A protostellar
outflow would likely expire less dramatically.  The ensuing
rarefaction and cavity infall would proceed with less voracity than in
these models and the deviation of the cavity size from the scaling
relation should be less than the prediction given by our models.
Therefore, the net momentum inferred from the size of an observed
cavity at $t>t_{off}$ using the scaling relation should be accurate to
within a factor of a few.  By $t \sim 40 \textrm{kyr}$, the radial
edges of both the jet and wide angle wind driven cavities have
decelerated to $< 1~\textrm{km s}^{-1}$.  At later time, a real fossil
cavity would be disrupted by the ambient turbulence of its parent
cloud.  The cavity would become observationally confused with the
ambient turbulence.  Figure 7 shows that the scaling relation provides
a prediction of the cavity volume that is accurate to within a factor
of a few for $t_{off}<t<40~\textrm{km s}^{-1}$.  Thus we conclude that
the scaling laws capture a reasonable approximation of the cavity
behavior and the conclusions of \cite{Quillen} relating to the
momentum injection, outflow dynamics and turbulent energy balance in
NGC 1333 are supported by our simulations.

\subsection{PV diagrams}
In this section we present synthetic Position Velocity (PV) diagrams
of our simulations.  In particular, we seek characteristics of fossil
outflows that are readily distinguished by observation.  PV diagrams
are commonly used to visualize observational data of molecular
outflows. Specifically, PV diagrams represent the distribution of
molecular gas at each velocity in the direction of the line of sight
at each position along a slit placed across a molecular outflow.  We
construct PV diagrams for each of the simulations presented here at
four angles of inclination to the plane of the sky, $0^\circ$,
$30^\circ$, $60^\circ$ and $90^\circ$.  The synthetic PV data
presented below have been processed through a Gaussian filter with
half-width $\sigma_x=0.0065~\textrm{pc}$ and $\sigma_v=0.0025 \times
~\textrm{(the maximum speed in the grid)}$.

\begin{figure}[!h]
\begin{center}
   \includegraphics[angle=-90,clip=true,width=0.333\textwidth]{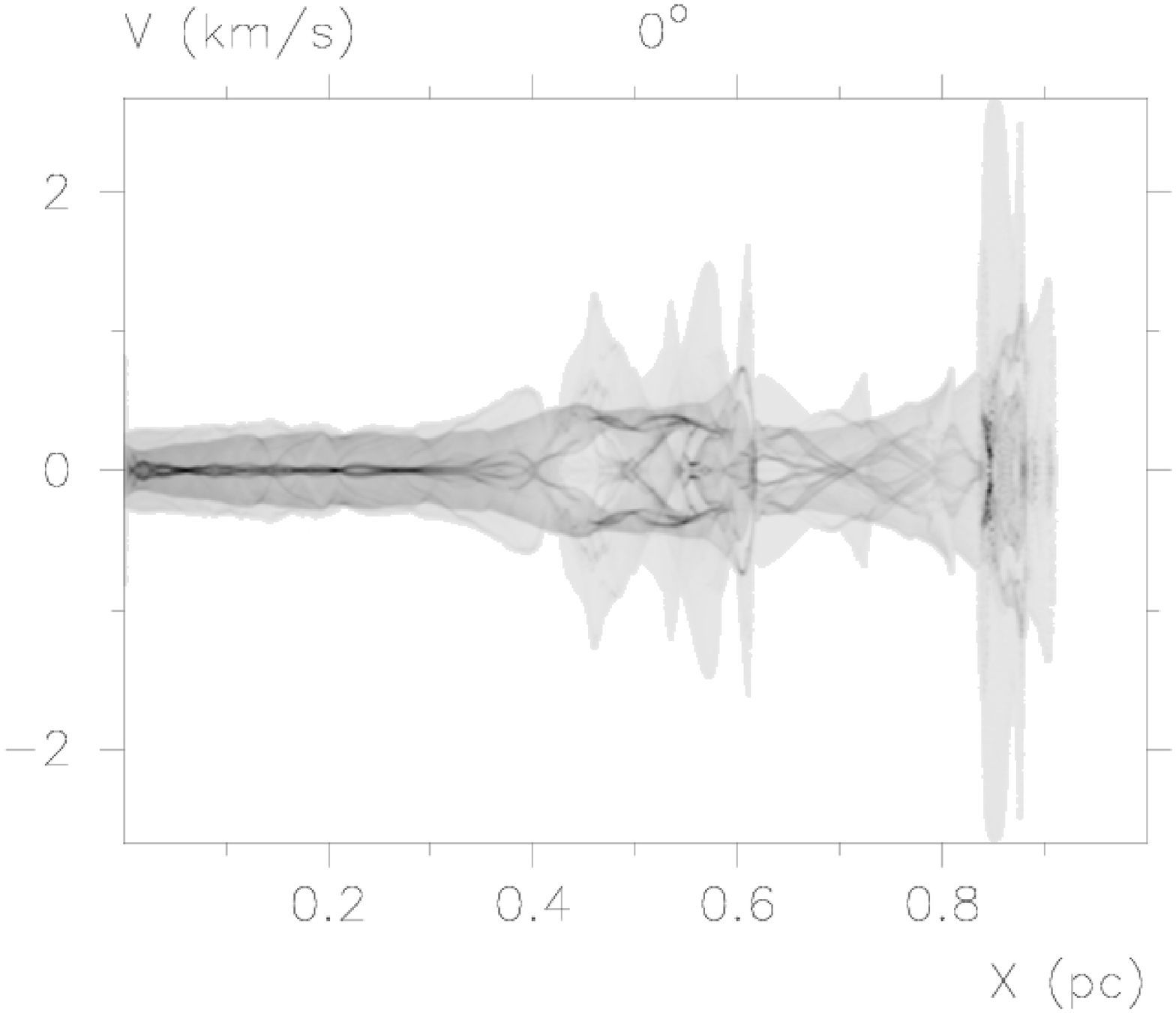}
   \includegraphics[angle=-90,clip=true,width=0.333\textwidth]{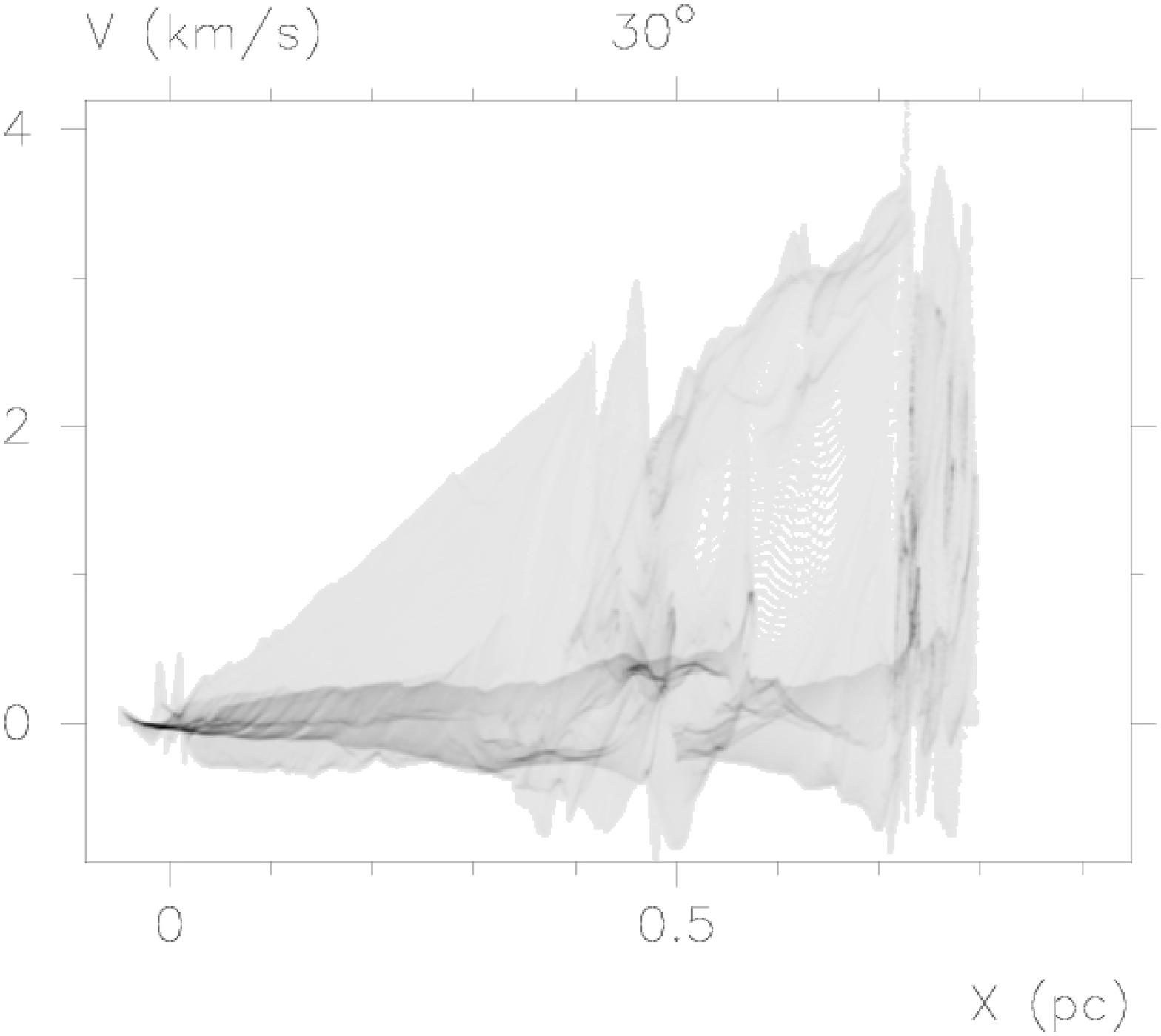}\\
   \includegraphics[angle=-90,clip=true,width=0.333\textwidth]{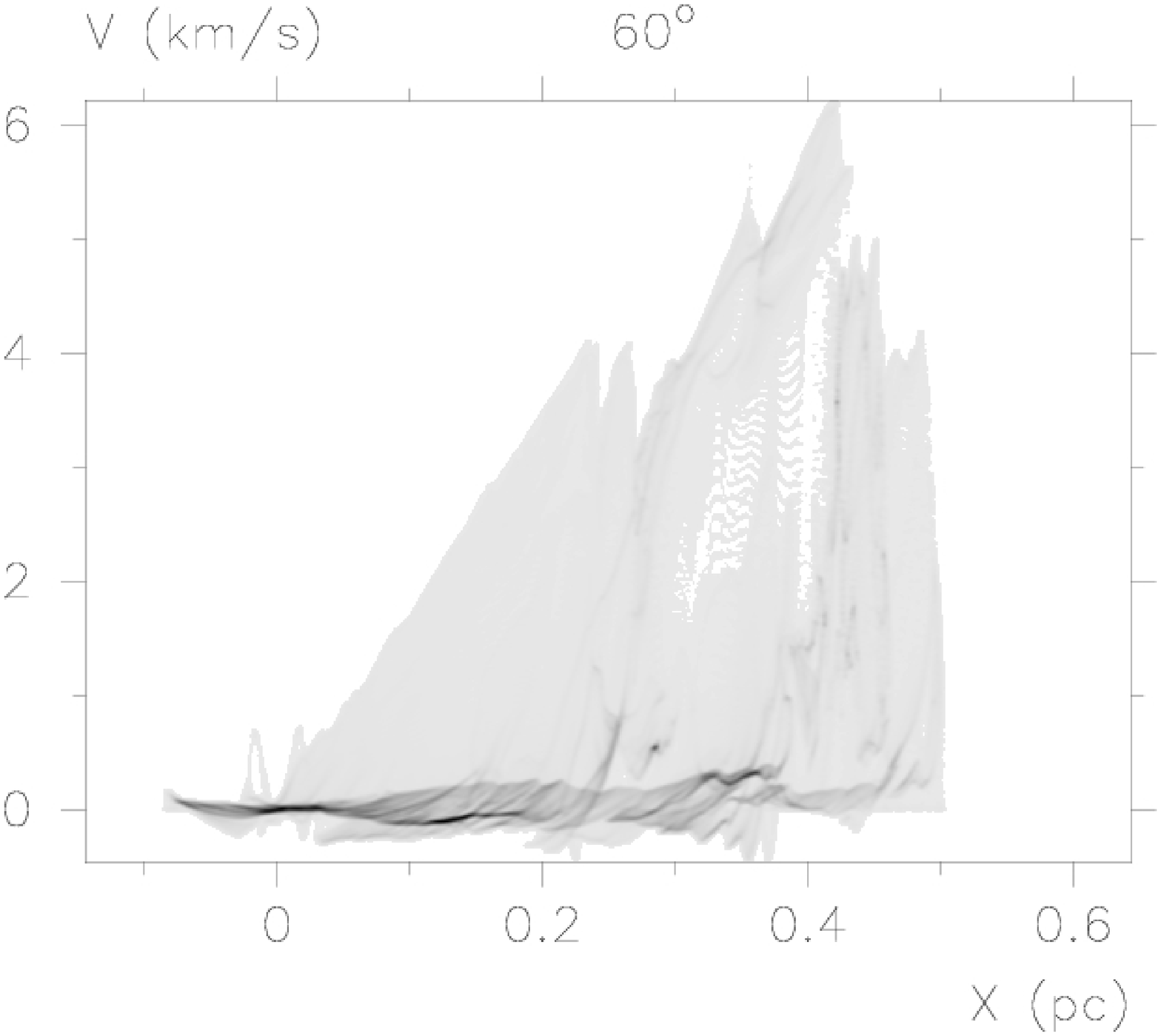}
   \includegraphics[angle=-90,clip=true,width=0.333\textwidth]{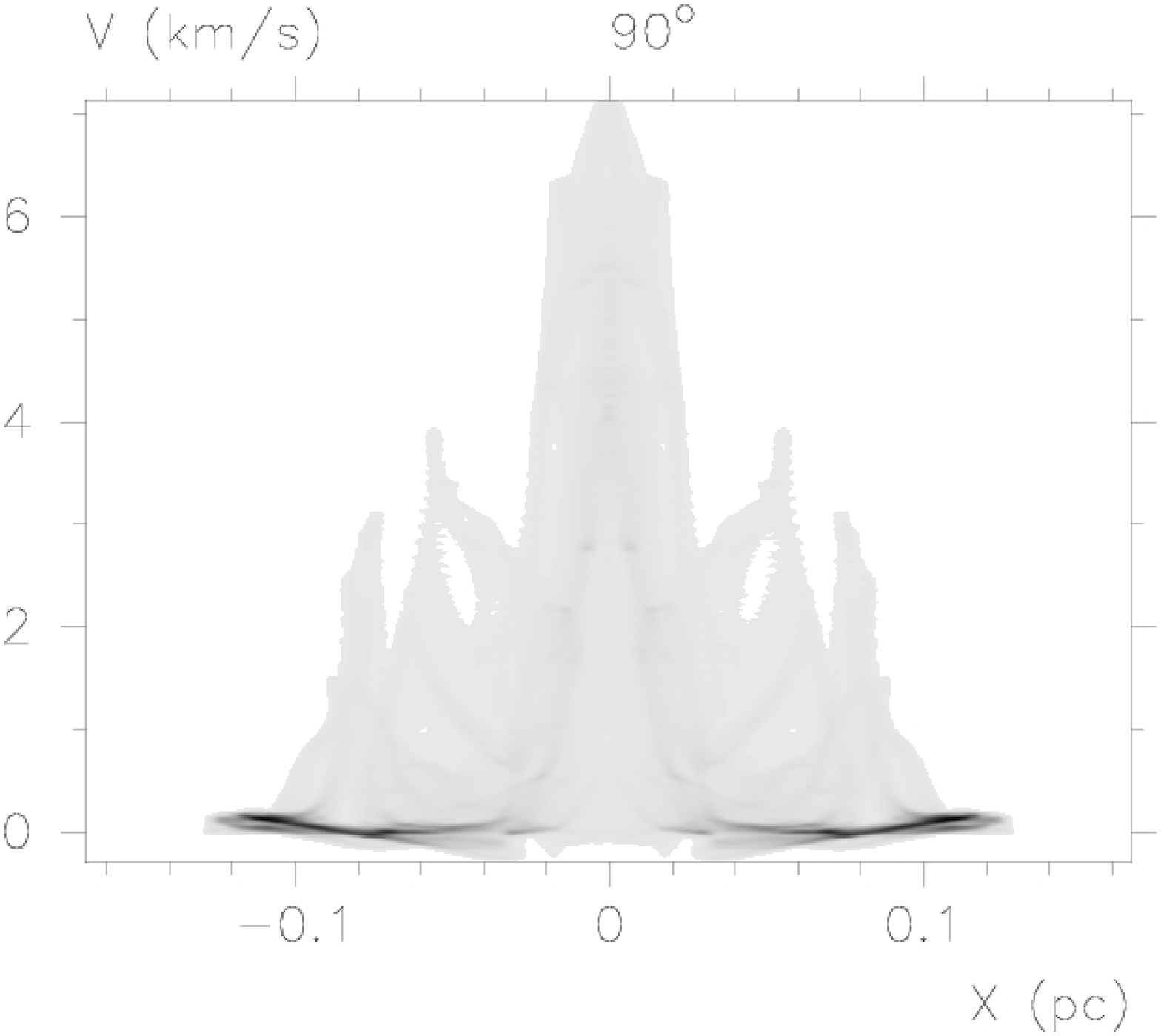}
\end{center}
\caption{ Position-velocity cuts of the molecular gas along the z-axis
of the fossil jet outflow $t=100~\textrm{kyr}$ (figure \ref{f1}, top)
with $0^\circ$ (top left), $30^\circ$ (top right), $60^\circ$ (bottom
left), and $90^\circ$ (bottom right) inclination to the plane of the
sky.  A Hubble law flow is exhibited in the $30^\circ$ and $60^\circ$
PV diagrams.
\label{f8}}
\end{figure}

In figure \ref{f8} synthetic PV diagrams of molecular gas in the
simulation domain have been taken along the z-axis at four angles of
inclination to the plane of the sky for the jet driven fossil cavity.
The densest gas along the outer wall of the cavity appears as a
horizontal bell-shaped distribution of gray scales when the fossil
cavity is oriented with its symmetry axis parallel to the plane of the
sky (upper left).  Note that gas expelled radially from the head of
the outflow subsequently decelerates as it flows backwards (in the
frame of the outflow head).  The gas that is closest to the outflow
source has decelerated the most.  At these positions gas appearing at
higher velocity than the cavity walls is backfilling from the inner
cavity wall into the cavity.  By the end of the simulation the speed
of the radial infall exceeds the radial expansion of the cavity walls.

At oblique orientations (figure \ref{f8}, upper right and lower left),
we identify a distinct right triangle-shaped feature.  The vertical
leg of the triangle represents the fastest moving gas at the head of
the outflow.  The horizontal leg appears due to the subsonic backfill
of the cavity walls.  The hypotenuse is the Hubble law flow formed
through velocity segregation as material closest to the head of the
flow propagates fastest and material closest to the source has
decelerated the most.

\begin{figure}[!h]
\begin{center}
   \includegraphics[angle=-90,clip=true,width=0.333\textwidth]{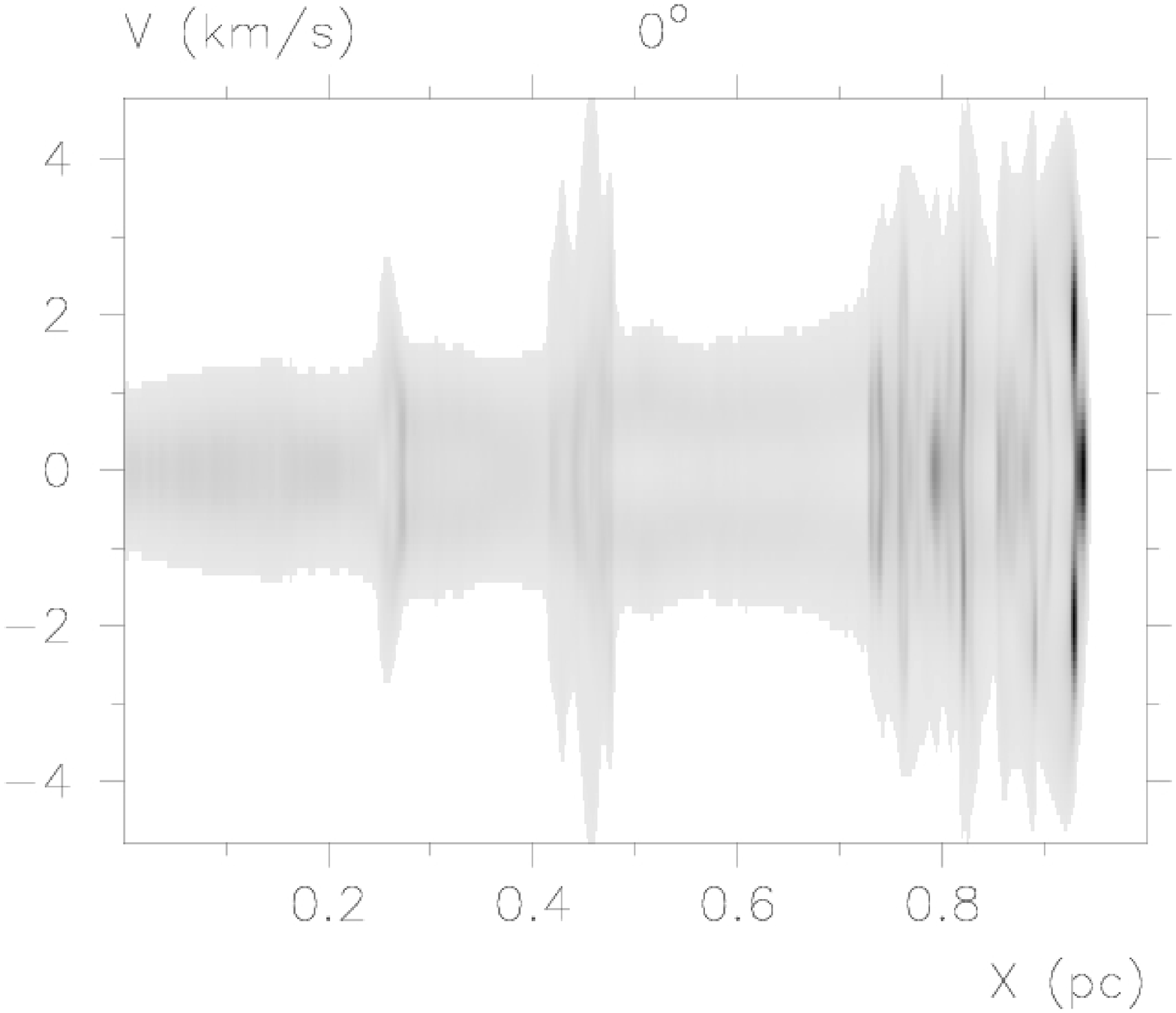}
   \includegraphics[angle=-90,clip=true,width=0.333\textwidth]{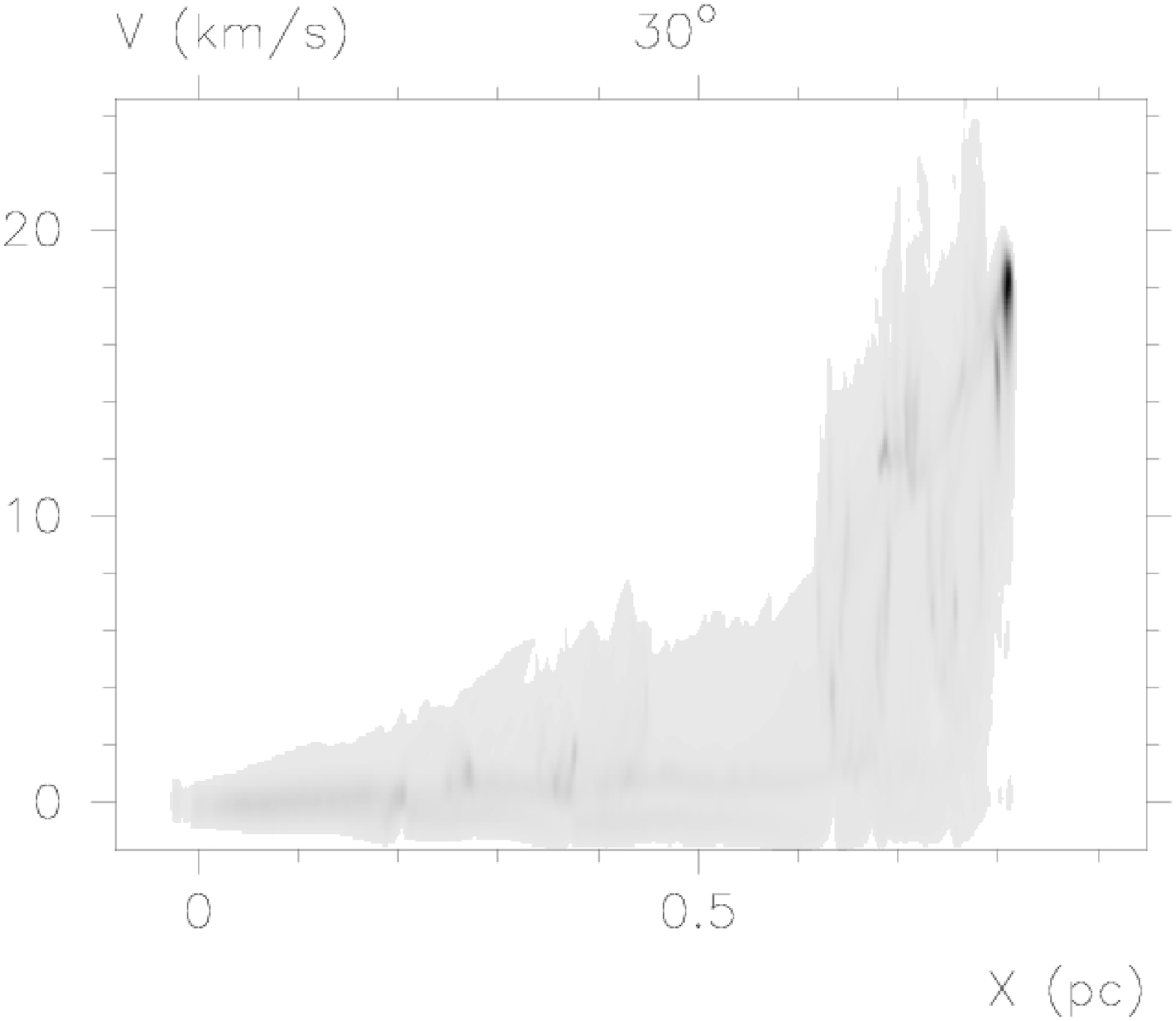}\\
   \includegraphics[angle=-90,clip=true,width=0.333\textwidth]{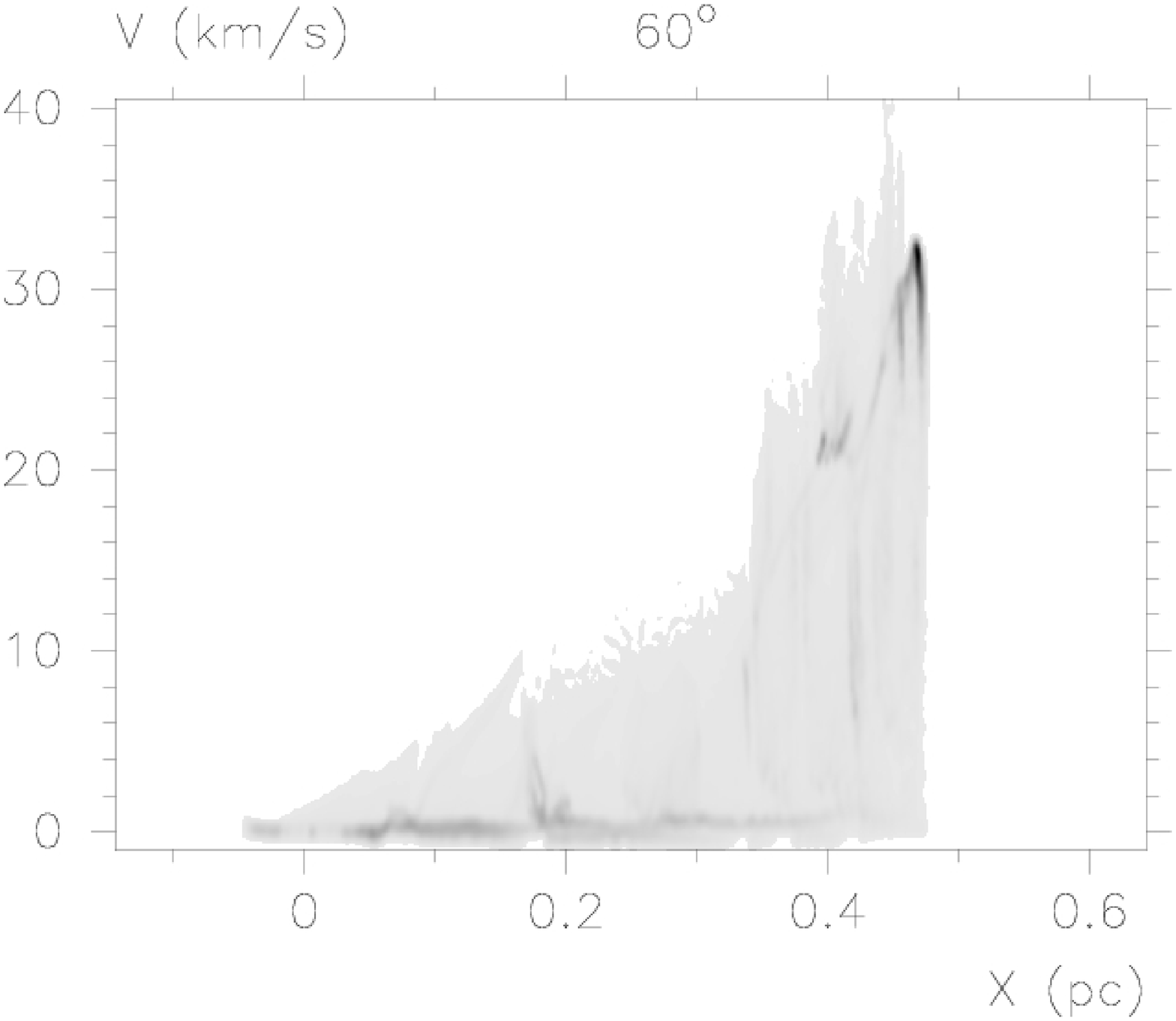}
   \includegraphics[angle=-90,clip=true,width=0.333\textwidth]{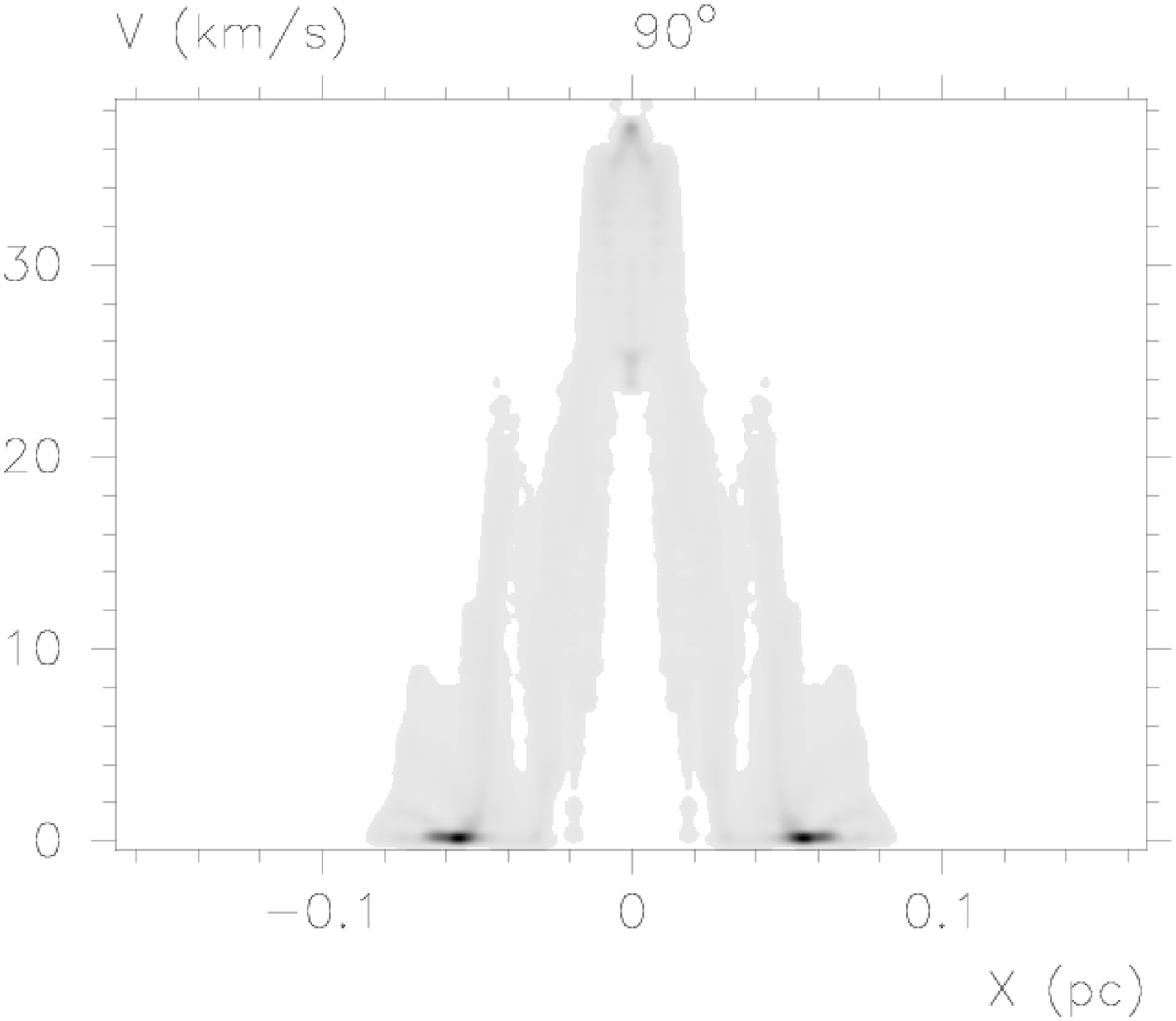}
\end{center}
\caption{ Position-velocity cuts of the molecular gas along the z-axis
 of the continuously driven jet outflow $t=34~\textrm{kyr}$ (figure
 \ref{f1}, bottom) with $0^\circ$ (top left), $30^\circ$ (top right),
 $60^\circ$ (bottom left), and $90^\circ$ (bottom right) inclination
 the plane of the sky.  No Hubble law flows are present here, though
 the spur structures described by \cite{lee01} are exhibited in the
 $30^\circ$ and $60^\circ$ PV diagrams.
\label{f9}}
\end{figure}

\begin{figure}[!h]
\begin{center}
   \includegraphics[angle=-90,clip=true,width=0.333\textwidth]{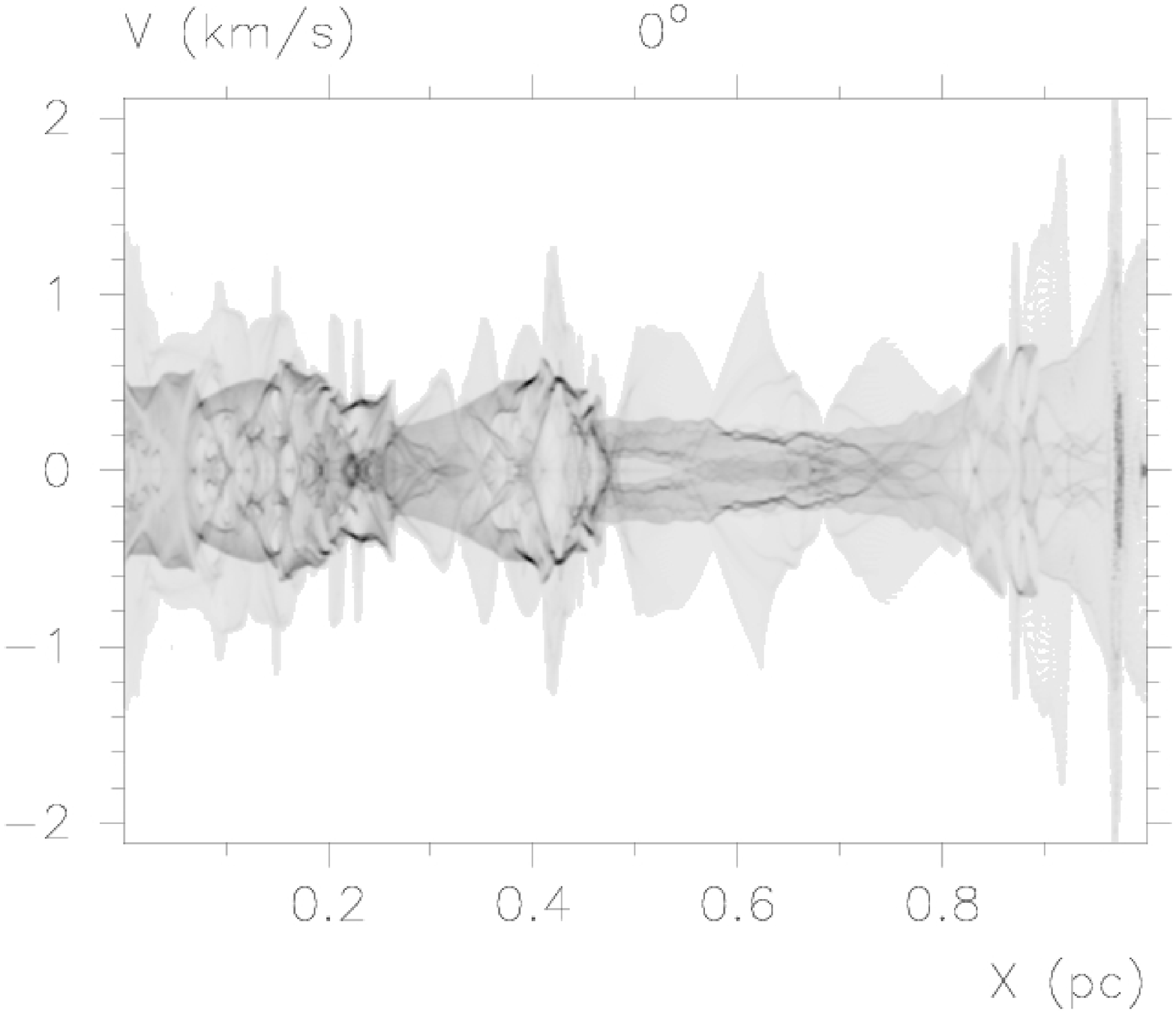}
   \includegraphics[angle=-90,clip=true,width=0.333\textwidth]{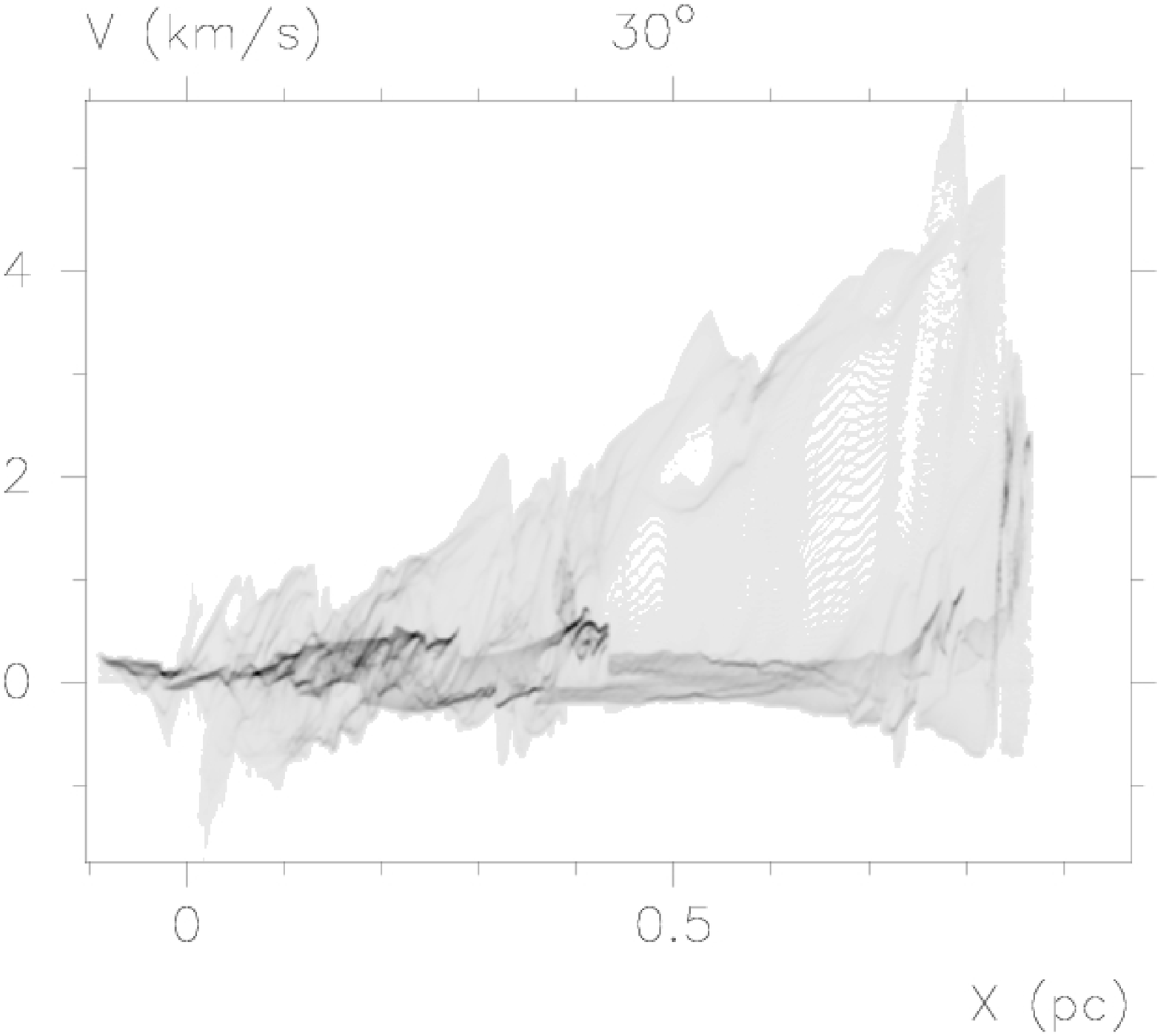}\\
   \includegraphics[angle=-90,clip=true,width=0.333\textwidth]{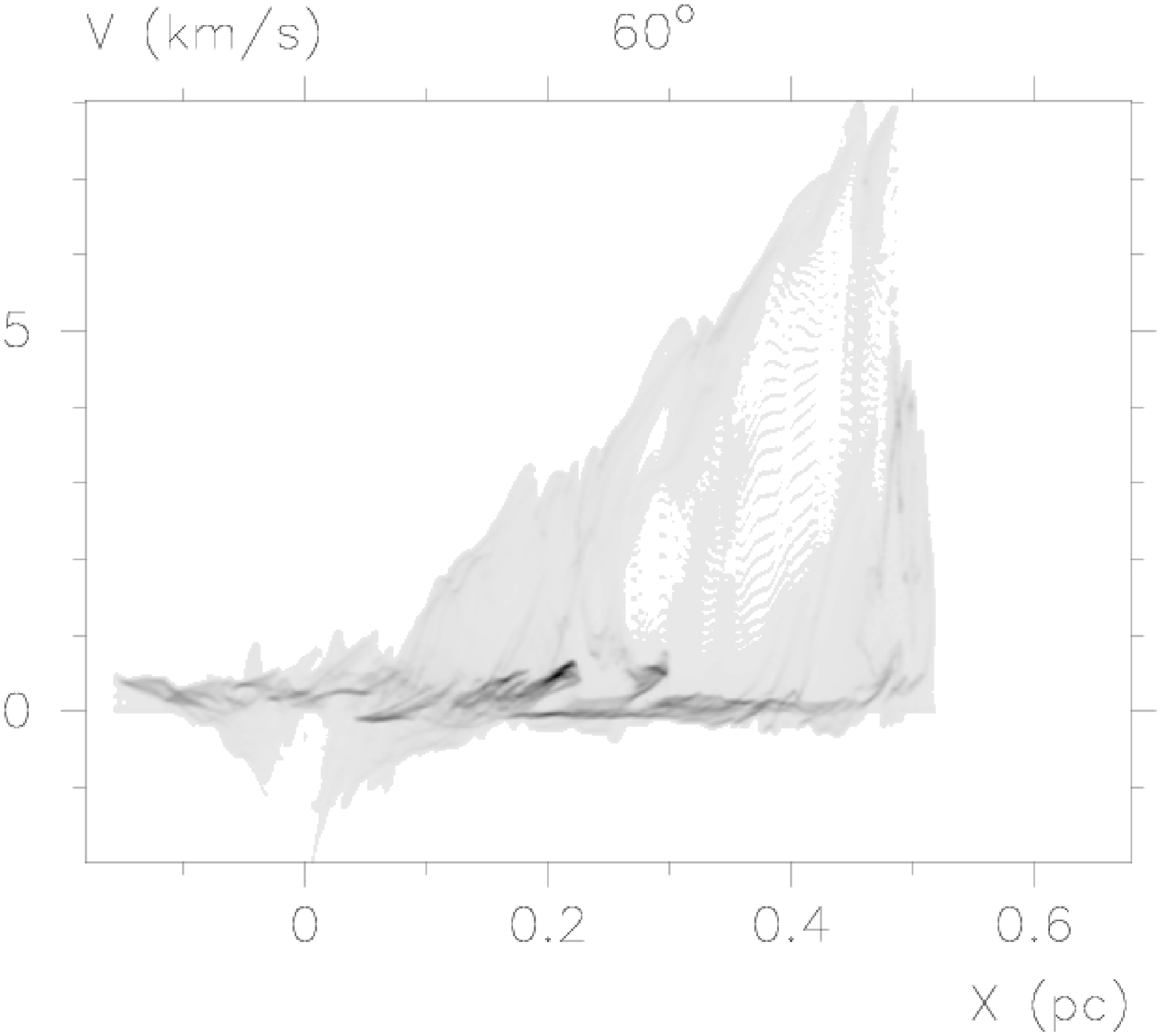}
   \includegraphics[angle=-90,clip=true,width=0.333\textwidth]{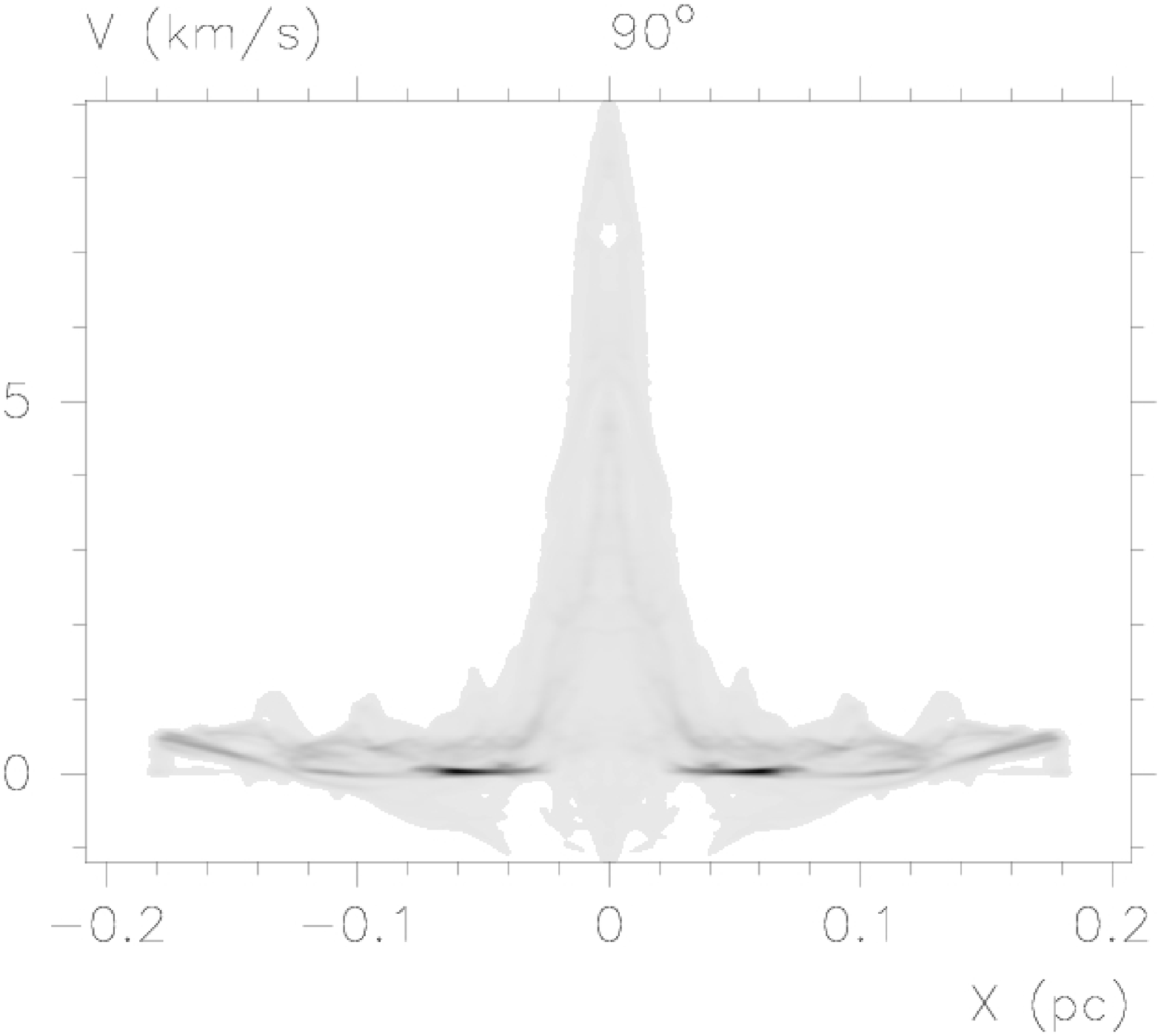}
\end{center}
\caption{ Position-velocity cuts of the molecular gas along the z-axis
 of the fossil wide angle wind outflow at $t=89~\textrm{kyr}$ (figure
 \ref{f2}, top) with $0^\circ$ (top left), $30^\circ$ (top right),
 $60^\circ$ (bottom left), and $90^\circ$ (bottom right) inclination
 to the plane of the sky.
\label{f10}}
\end{figure}

\begin{figure}[!h]
\begin{center}
   \includegraphics[angle=-90,clip=true,width=0.333\textwidth]{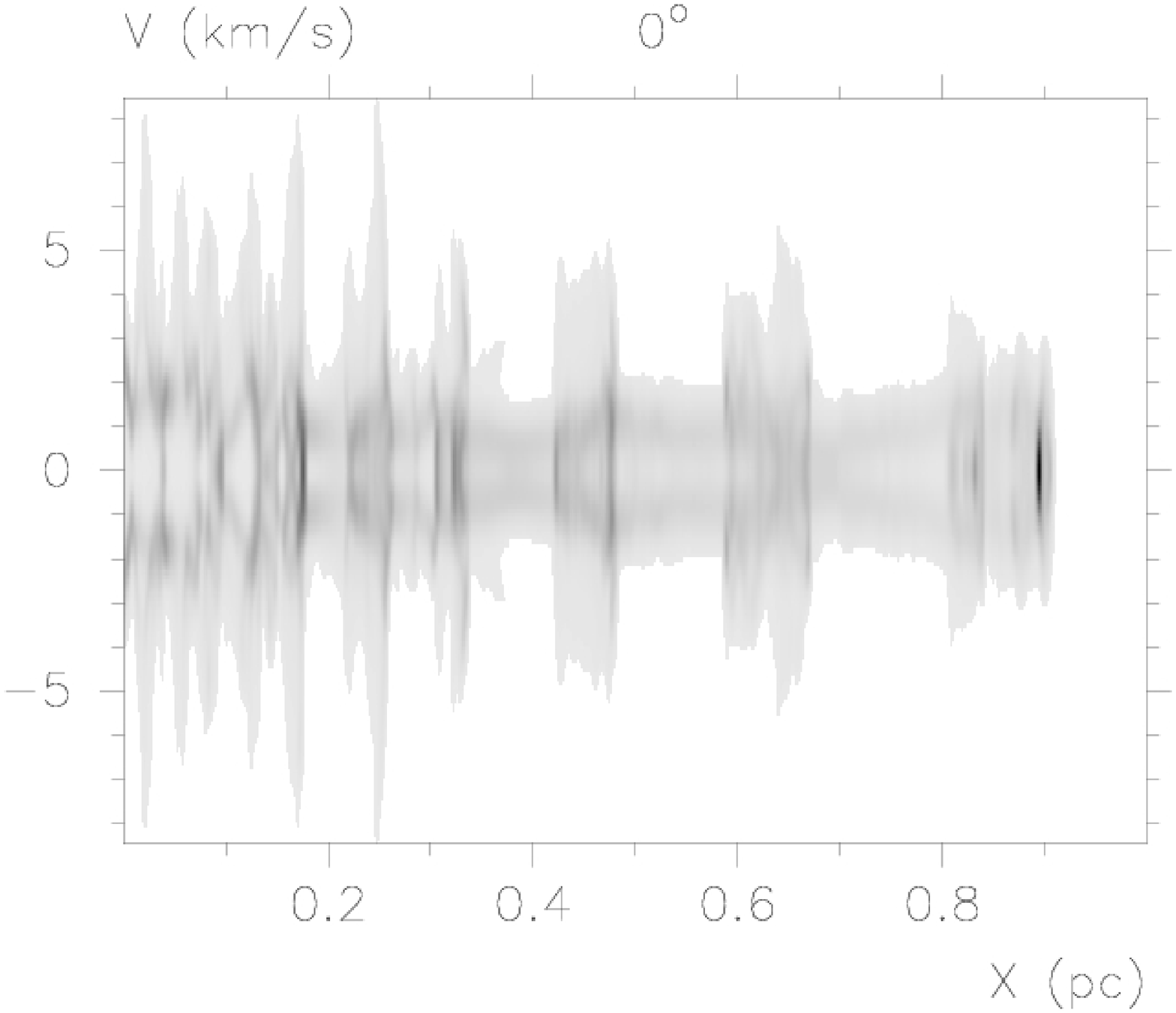}
   \includegraphics[angle=-90,clip=true,width=0.333\textwidth]{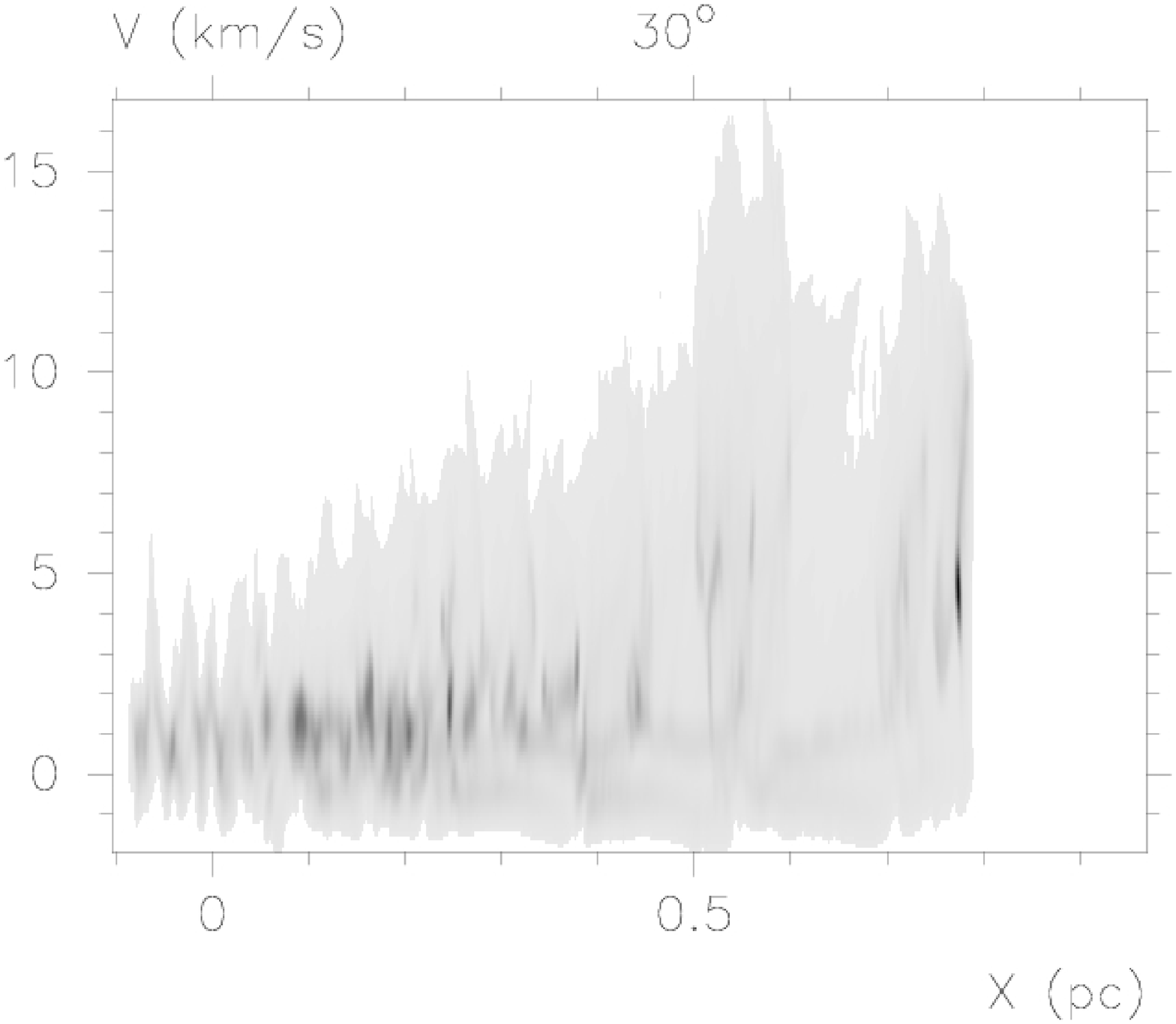}\\
   \includegraphics[angle=-90,clip=true,width=0.333\textwidth]{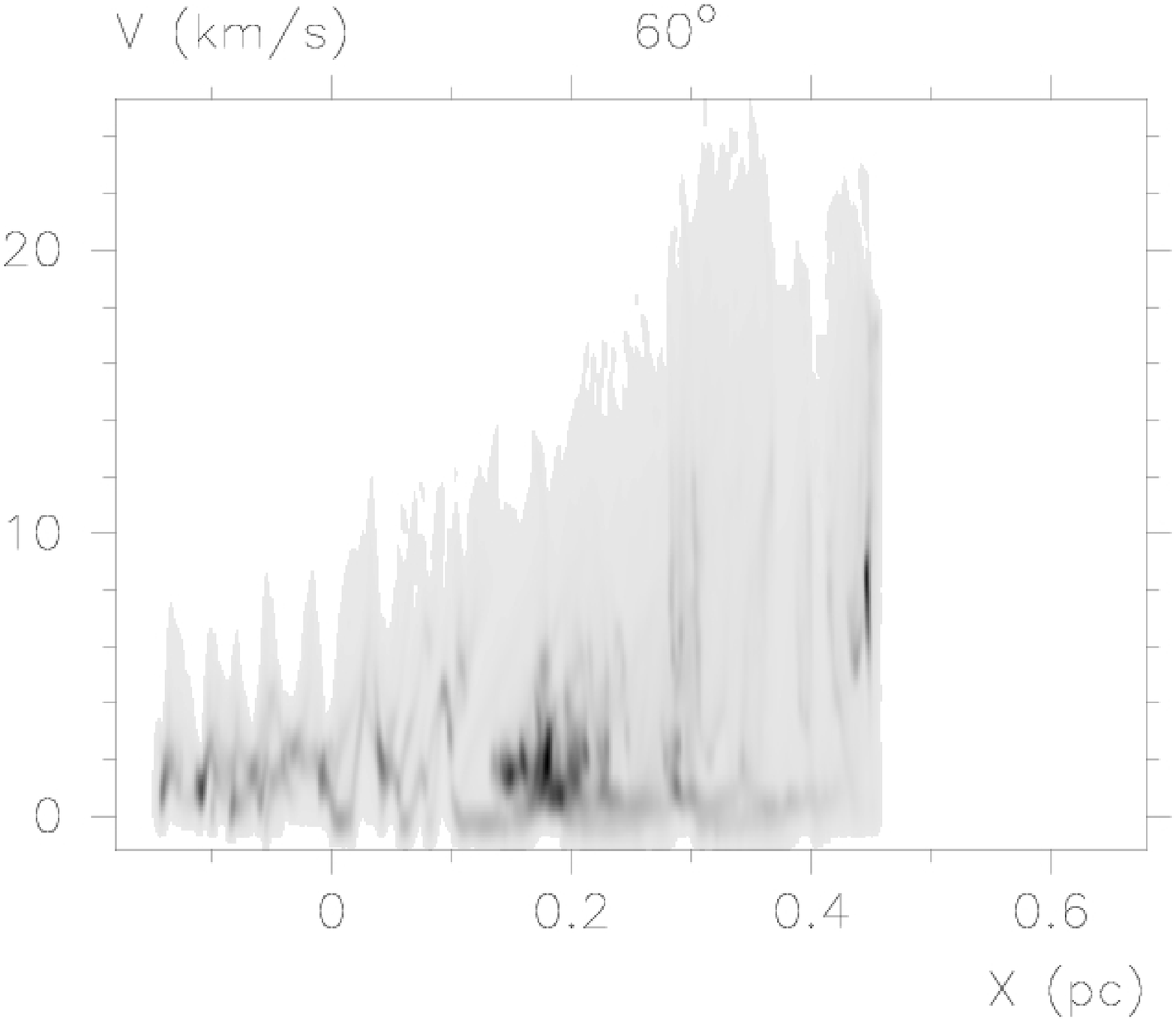}
   \includegraphics[angle=-90,clip=true,width=0.333\textwidth]{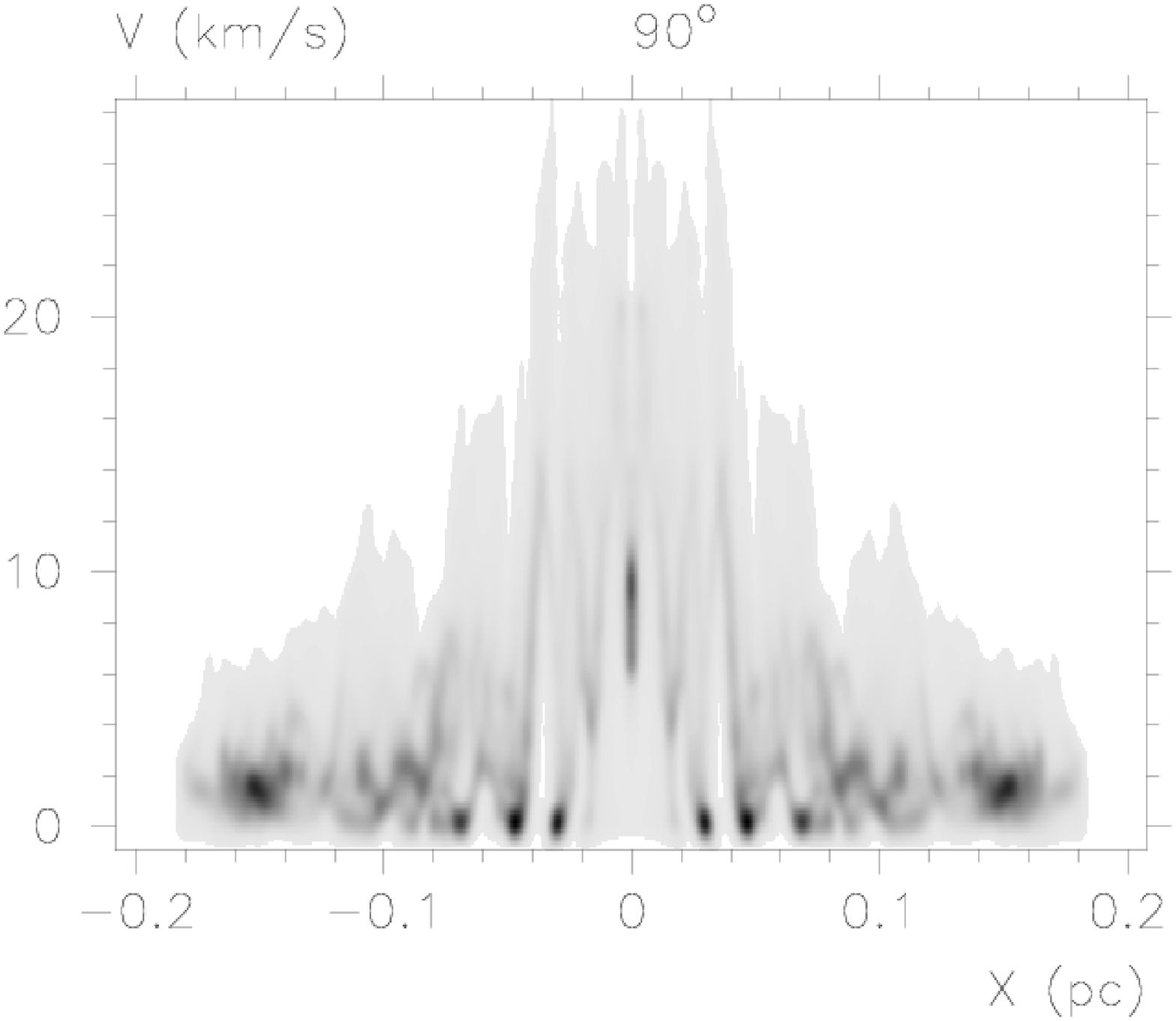}
\end{center}
\caption{ Position-velocity cuts of the molecular gas along the z-axis
 of the continuously driven wide angle wind outflow at $t=33~\textrm{kyr}$ (figure \ref{f2},
 bottom) with $0^\circ$ (top left), $30^\circ$ (top right), $60^\circ$
 (bottom left), and $90^\circ$ (bottom right) inclination to the plane of the sky.
\label{f11}}
\end{figure}

When the line of sight with the outflow axis is oriented perpendicular
to the plane of the sky (lower right) only the axial velocity
contributes to the PV diagram.  Now the fastest moving material
appears within the observed central spike and the most decelerated
material appears furthest from the central source.
% Figure 8
%\clearpage
%\clearpage
We can contrast the PV distribution of the jet driven fossil cavity
with its continuously driven counterpart (figure \ref{f9}).  Again,
the densest gas delineates a bell shape when the outflow axis is
oriented parallel to the plane of the sky.  In this case there is no
stationary material in contrast to the fossil cavity.  Because the
outflow is driven from the boundary of the grid is initially atomic
and hence not revealed in the PV data and the continuous driving does
not allow backfilling into the cavity, there is no transition between
the radial expansion and infall within the outflow structure and hence
no stationary gas.  Also, in contrast to the fossil cavity at oblique
inclinations, the velocity of gas follows a faster than linear
increase in velocity with distance from the source.  Gas that is
radially expelled from the head of the outflow is highly supersonic
with $|v| \sim 30~\textrm{km s}^{-1}$.  The propagation of the head of
the fossil cavity is nearly an order of magnitude lower.  The
continuously driven jet is subject to greater radiative energy loss
and hence a lower cocoon pressure than the fossil jet.  The greater
loss of this pressure support causes radially expanding gas to undergo
a more rapid deceleration than in the fossil cavity case, yielding the
faster increase in velocity from the outflow source.
% Figure 9
%\clearpage
%\clearpage
Most of the features of the PV data for jet driven fossil cavities
applies to the PV data of the wide angle wind driven cavity (figure
\ref{f10}).  One notable difference is that the densest gas along the
cavity wall, delineated by a bell shape at zero inclination to the
plane of the sky, retains some radial expansion close to the outflow
source.  The wide angle wind applies ram pressure driving to the
entire perimeter of the cavity walls rather than solely at the head of
the outflow as in the jet driven case.  This prevents the stall of
radial expansion of the base of the outflow.
% Figure 10
%\clearpage
%\clearpage
Similarly, the features present in the PV data for continuously driven
jets apply also to the continuously driven wide angle wind (figure
\ref{f11}).  In this case, however, the radial expansion of the
outflow near the source exceeds the radial expansion at the tip of the
outflow.  This is due to the isotropic ram pressure of the driving
wind that prohibits the stall of the cavity expansion near the base of
the source.
% Figure 11
%\clearpage
%\clearpage
In each configuration, jet, wide angle, fossil and driven, the
condensed gas at the head of the cavity has retained most of the
momentum injected by the wind.  This appears as a central spike in the
PV diagrams where the outflow axis is oriented directly toward the
observer.  The wide angle wind cases reveal greater rates of radial
expansion near the base of the outflow than their jet driven
counterparts.  Because the wide angle winds used in these simulations
produce a relatively collimated outflow structure, this is only easily
distinguished when the outflow axis is oriented along the plane of the
sky.  In the simulations of continuously driven winds (figures
\ref{f9} \& \ref{f11}), most of the gas entrained in the outflow is
concentrated at velocities that follow a slightly faster than linear
increase in velocity with position from the driving source at oblique
inclinations.  This result was also noted in the jet driven outflow
models of \cite{lee01}.  Molecular gas that has backfilled into the
cavity after the driving source has expired appears at each position
in the PV diagram at low and intermediate velocity.

%%\clearpage
\section{Conclusions}
In this paper we have presented AMR simulations of molecular outflows
driven into an ambient molecular cloud.  We have compared simulations
of a fossil cavity driven by a previously active jet, and that of a
continuously driven jet.  We also have compared simulations of a
fossil cavity driven by a previously active wide angle wind to that of
a continuously driven wind.  Previous numerical simulations of
molecular outflow have focused only on outflows driven by active
sources.  We find that the fossil cavities reach similar length scales
as their continuously driven counterparts at much later times.  Their
aspect ratios approach a constant value, wheres those of the
continuously driven outflows continue to increase.  Hence the fossil
cavities are wider than their continuously driven counterparts.  We
find that the density contrast (the ratio of the density outside the
cavity to that inside) drops with time, due to backfill of wind-swept
ambient gas into the cavity.  At later times fossil cavities can have
low density contrasts of order unity.

We have shown that a scaling law for self-similar flow derived from
momentum conservation roughly predicts the volume of the fossil
cavity.  The fossil cavities expand somewhat faster than predicted
from the scaling law.  The success of the scaling law implies that
observed sizes and velocities of cavities can be used to estimate the
total momentum input required to create them.  This point is crucial
when using outflows as sources in energy budget arguments for
re-energizing molecular cloud turbulent motions.

We find that synthetic position velocity plots of the fossil cavities
exhibit Hubble law expansion patterns at intermediate viewing angles,
whereas both fossil and active outflows contain spur patterns caused
by the head of a bow shock. Our simulations of continuously driven
outflows lack Hubble law patterns in the PV plots.  It is tempting to
consider the possibility that cavities that exhibit Hubble flows are
fossil cavities rather than continuously driven outflows.  However,
previous simulations that have enhanced entrainment have also
succeeded in showing Hubble laws.  Furthermore some observed Hubble
law flows have high velocities and so are likely to be actively driven
(see figure 2 of \citealt{arce01}).

The interpretation of slowly expanding cavities in NGC 1333 as the
relics of past outflow activity is supported by our simulations.  The
simulated fossil outflow cavities have a lifetime $\sim 10\times$ that
of the duration of their driving source.  Therefore, we expect fossil
cavities to appear with $10 \times$ the frequency of active outflows
in molecular star forming regions.  \cite{Quillen} arrived at a
similar estimate from statistical considerations of the frequency of
cavities they observed in NGC 1333 compared to the frequency of active
outflows. 

The outflow swept cavities that remain as relics of previous YSO outflow
activity provide a link between outflow activity and the turbulent
support of some molecular clouds.  Fossil outflows retain speeds above
the turbulent velocity of their parent cloud on a timescale greater
than the lifetime of the driving source.  Therefore, the fossil relic
stage of an outflow's lifetime will result in the entrainment of a
considerable amount of ambient molecular material in addition to that
which was swept up during the lifetime of the driving source.  As
ambient gas entrained into the cavity flow, the expansion of cavities
will decelerate to speeds comparable to the turbulent speed of the
ambient could.  The coupling of the momentum carried by cavities will
couple to the turbulent motions in the could via their disruption by
the preexisting turbulent media into which they are injected, by the
slow collision and overlap of two or more cavities or both.  Future
work should study the details of the mechanisms by which fossil
cavities can re-energize a turbulent media through such coupling.

\acknowledgments We acknowledge support for this work from the Jet
Propulsion Laboratory Spitzer Space Telescope theory grant 051080-001,
Hubble Space Telescope theory grant 050292-001, National Science
Foundation grants AST-0507519, AST-0406799, AST 00-98442 \&
AST-0406823, DOE grant DE-F03-02NA00057, the National Aeronautics and
Space Administration grants ATP04-0000-0016 \& NNG04GM12G issued
through the Origins of Solar Systems Program.  We also adknowledge the
computational resources provided by University of Rochester
Information Technology Services and the Laboratory for Laser
Energetics.

\end{document}